\documentclass[conference, compsoc]{IEEEtran}

\usepackage[utf8]{inputenc}
\usepackage[T1]{fontenc}
\usepackage{amsmath,amssymb,amsfonts,amsthm}
\usepackage{graphicx}
\usepackage{booktabs}
\usepackage{multirow}
\usepackage[table]{xcolor}
\usepackage{xurl}
\usepackage[breaklinks]{hyperref}
\usepackage{enumitem}
\usepackage{pifont}
\usepackage{balance}
\usepackage{algorithm}
\usepackage{algorithmic}
\usepackage{threeparttable}
\usepackage{tabularx}
\usepackage{makecell}
\usepackage{xspace}
\usepackage{xcolor}

\usepackage{array}
\newcolumntype{L}[1]{>{\raggedright\arraybackslash}m{#1}}
\newcolumntype{C}[1]{>{\centering\arraybackslash}m{#1}}
\newcolumntype{R}[1]{>{\raggedleft\arraybackslash}m{#1}}
\definecolor{pone}{HTML}{F3E9D6} 
\definecolor{ptwo}{HTML}{F1DFC1} 

\theoremstyle{plain}
\newtheorem{definition}{Definition}

\newcommand{\cmark}{\ding{51}}
\newcommand{\xmark}{\ding{55}}
\newcommand{\pmark}{{\color{orange}$\sim$}}
\newcommand{\para}[1]{\smallskip\noindent\textbf{#1.}}

\hypersetup{
  colorlinks=true,
  linkcolor=blue!70!black,
  citecolor=blue!70!black,
  urlcolor=blue!70!black
}

\newcommand{\openclaw}{\raisebox{-0.3em}{\includegraphics[height=1.5em]{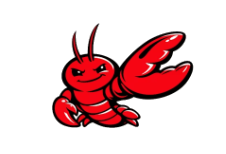}}}

\begin{document}

\title{SoK: Agentic Skills — Beyond Tool Use in LLM Agents}

\author{
  \IEEEauthorblockN{Yanna Jiang$^1$, Delong Li$^1$, Haiyu Deng$^1$, Baihe Ma$^1$,  Xu Wang$^1$, Qin Wang$^{1,2}$,  Guangsheng Yu$^1$}
  \IEEEauthorblockA{$^1$University of Technology Sydney $|$ $^2$CSIRO Data61}
}
\maketitle

\begin{abstract}
Agentic systems increasingly rely on reusable procedural capabilities, \textit{a.k.a., agentic skills},  to execute long-horizon workflows reliably. These capabilities are callable modules that package procedural knowledge with explicit applicability conditions, execution policies, termination criteria, and reusable interfaces. Unlike one-off plans or atomic tool calls, skills operate (and often do well) across tasks.

This paper maps the skill layer across the full lifecycle (discovery, practice, distillation, storage, composition, evaluation, and update) and introduces two complementary taxonomies. The first is a system-level set of \textbf{seven design patterns} capturing how skills are packaged and executed in practice, from metadata-driven progressive disclosure and executable code skills to self-evolving libraries and marketplace distribution. The second is an orthogonal \textbf{representation $\times$ scope} taxonomy describing what skills \emph{are} (natural language, code, policy, hybrid) and what environments they operate over (web, OS, software engineering, robotics). 

We analyze the security and governance implications of skill-based agents, covering supply-chain risks, prompt injection via skill payloads, and trust-tiered execution, grounded by a case study of the ClawHavoc campaign in which nearly 1{,}200 malicious skills infiltrated a major agent marketplace, exfiltrating API keys, cryptocurrency wallets, and browser credentials at scale. We further survey deterministic evaluation approaches, anchored by recent benchmark evidence that curated skills can substantially improve agent success rates while self-generated skills may degrade them. We conclude with open challenges toward robust, verifiable, and certifiable skills for real-world autonomous agents.

\end{abstract}

\section{Introduction}
\label{sec:intro}

Large language model (LLM) agents have advanced rapidly from single-turn question answering to \textit{multi-step} autonomous systems that browse the web~\cite{zhou2024webarena}, write/debug software~\cite{yang2024sweagent,ji2025measuring}, orchestrate tools in sequence~\cite{shen2023hugginggpt,xie2024can}, and collaborate as multi-agent teams~\cite{hong2024metagpt,wu2024autogen}. 
Yet a fundamental inefficiency persists: each new task forces the agent to re-derive an execution strategy from scratch. A coding agent that has successfully debugged a null-pointer exception a hundred times still approaches the hundred-and-first as if it were novel. The procedural knowledge gained from experience disappears at the end of every context window.

This observation motivates the central abstraction of this paper: the \emph{agentic skill}s. We define a skill as a reusable, callable module that encapsulates a sequence of actions or policies enabling an agent to achieve a class of goals under recurring conditions. Skills differ from tools (atomic primitives with fixed interfaces), plans (one-time reasoning scaffolds), and episodic memories (stored observations) in that they are simultaneously \emph{executable}, \emph{reusable}, and \emph{governable}. A skill carries its own applicability conditions, termination criteria, and callable interface, making it a first-class unit of procedural knowledge.

The notion is not new in isolation. Cognitive architectures such as ACT-R~\cite{anderson2004actr} and Soar~\cite{laird2012soar} formalized procedural memory decades ago. Reinforcement learning (RL) has long studied option frameworks and hierarchical policies~\cite{sutton1999options}. 
What \emph{is} new is the convergence of these ideas in the LLM agent~\cite{zhanglandscape,qiantoolrl}. Skills manifest in forms ranging from natural-language playbooks and executable Python scripts to marketplace-distributed plugins. The diversity of representations calls for systematization.

Existing surveys cover LLM agents broadly~\cite{gaosurvey,ma2026safety,wang2024survey_agents,shahriar2025survey,huang2024understanding,guo2024large,yehudai2025survey}, focus on tool use~\cite{fan2026information,qin2024toollearning,schick2024toolformer}, or address multi-agent coordination~\cite{guo2024multiagent}. None adopt a \textit{skill-centric} lens to trace the full lifecycle from acquisition to governance. We fill this gap.

\para{Contributions} This Systematization of Knowledge (SoK) offers six contributions:
\begin{itemize}[leftmargin=*,nosep]
  \item a \textit{unified definition} of agentic skills (\S\ref{sec:definition}), formalized as $S = (C, \pi, T, R)$, with precise boundary conditions separating skills from tools, plans, and memory.
  \item a \textit{skill lifecycle model} (\S\ref{sec:lifecycle}) mapping the stages from discovery through evaluation and update, with a summary linking representative systems to lifecycle stages.
  \item a \textit{seven-pattern design taxonomy} (\S\ref{sec:patterns}) for how skills are packaged, loaded, and executed in real systems.
  \item an orthogonal \textit{representation $\times$ scope taxonomy} (\S\ref{sec:taxonomy}) describing what skills are and what environments they act over, integrated with the seven patterns.
  \item a \textit{security and governance analysis} (\S\ref{sec:security}) covering threat models, trust tiers, a pattern-specific risk matrix, and an anchor case study of the ClawHavoc marketplace supply-chain attack.
  \item an \textit{evaluation framework}  (\S\ref{sec:evaluation}) with metrics, benchmark mapping, and an anchor case study demonstrating that curated skills outperform self-generated ones.
\end{itemize}

\para{Reading map} \S\ref{sec:definition} defines the core abstraction. \S\ref{sec:methodology} describes our systematic methodology. \S\ref{sec:lifecycle} introduces the lifecycle model. \S\ref{sec:patterns} presents the seven design patterns and the representation$\times$scope taxonomy. \S\ref{sec:acquisition} covers skill acquisition and composition. \S\ref{sec:security} analyzes security and governance, including the ClawHavoc case study. \S\ref{sec:evaluation} surveys evaluation. \S\ref{sec:discussion} discusses cross-cutting observations and limitations. \S\ref{sec:openproblems} outlines open challenges. \S\ref{sec:conclusion} concludes our work.

\section{What Is an Agentic Skill?}
\label{sec:definition}

\subsection{Formal Definition}
\label{sec:def:formal}

We ground the concept of an agentic skill in a four-tuple formalization that captures the essential properties distinguishing skills from related abstractions.

\begin{definition}[Agentic skills]
Let an agent interact with environment $E$ via action space $A$, observation space $O$, and goal space $G$. Let $H = (o_1, a_1, \ldots, o_{t-1}, a_{t-1})$ denote the interaction history up to the current step. An \emph{agentic skill} is a tuple
\begin{equation}
  S = (C, \pi, T, R)
  \label{eq:skill_def}
\end{equation}
where:
\begin{itemize}[nosep,leftmargin=1em]
  \item $C: O \times G \rightarrow \{0,1\}$ is the \textbf{applicability condition}, a predicate over observations and the agent's current goal that determines whether the skill is appropriate for the current context;
  \item $\pi: O \times H \rightarrow A \cup \Sigma$ is the \textbf{executable policy}, a mapping from observations and interaction history to actions or skill invocations from the skill library $\Sigma$, which may be implemented as natural-language instructions, executable code, a learned controller, or a hybrid thereof. When $\pi$ selects a skill $s \in \Sigma$ rather than a primitive action $a \in A$, hierarchical composition arises (\S\ref{sec:composition:hierarchical}), mirroring the option-subroutine structure in the RL options framework~\cite{sutton1999options};
  \item $T: O \times H \times G \rightarrow \{0,1\}$ is the \textbf{termination condition}, specifying when the skill has completed (successfully or not) relative to the current goal;
  \item $R = (\textit{name}, \textit{params}, \textit{returns})$ is the \textbf{reusable callable interface}, a metadata and contract component specifying the skill's callable signature (name, parameter schema, return type) for programmatic invocation by the agent, other skills, or external orchestrators.
\end{itemize}
$G$ may be encoded within $O$ (e.g., as a task prompt in the observation) or passed as an explicit parameter; we make it explicit here for clarity. Implementations often compute soft applicability scores $C: O \times G \rightarrow [0,1]$ and apply a threshold; we present the binary form as a simplifying convention that captures the essential gating logic. In orchestrator-managed architectures, $C$ and $T$ may be externally provided rather than skill-internal; the 4-tuple then describes the logical interface regardless of where each function is implemented.
\end{definition}

We argue these four components form a useful minimal schema that captures the properties distinguishing skills from related abstractions. Removing $C$ yields a policy that cannot self-select; removing $T$ produces a policy that cannot compose (callers do not know when to resume); removing $R$ yields internal knowledge that cannot be invoked programmatically; and removing $\pi$ leaves metadata without executability. The formalization is deliberately representation-agnostic: $\pi$ can be a prompt template, a Python function, a reinforcement-learning policy, or a combination.

\begin{figure}[t]
  \centering
  \includegraphics[width=\columnwidth]{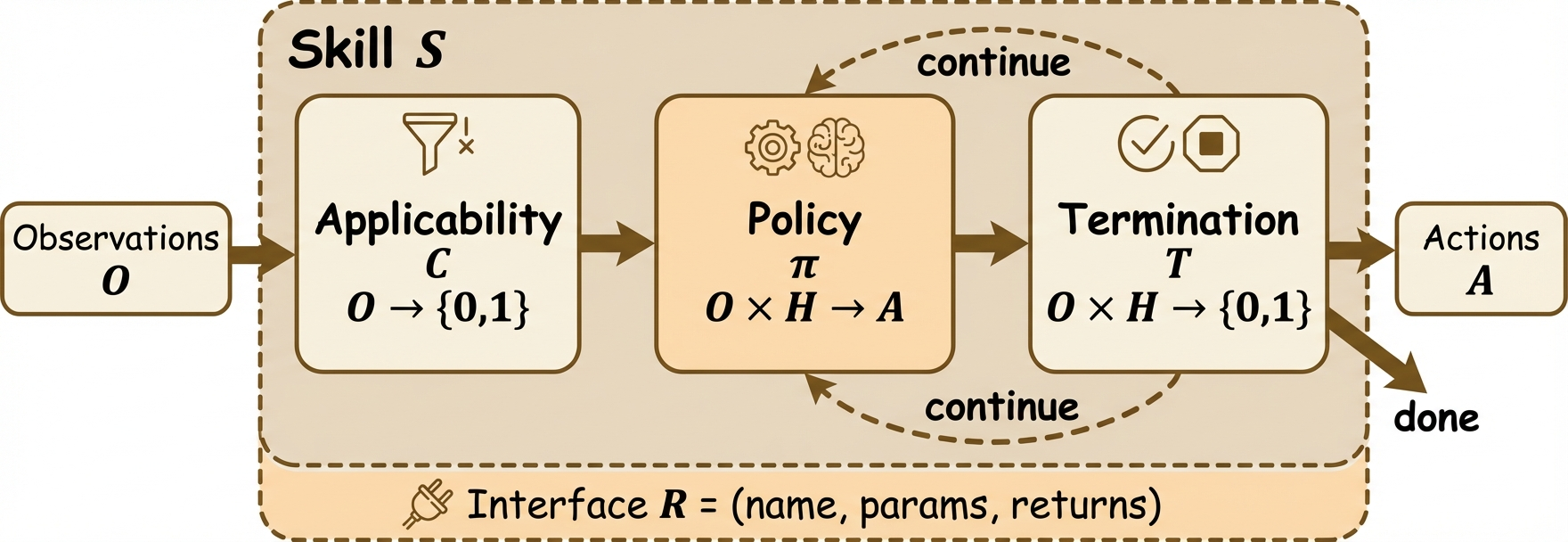}
  \caption{Internal anatomy of an agentic skill. Observations~$O$ enter the applicability gate~$C$; the policy~$\pi$ produces actions~$A$; the termination condition~$T$ determines whether to continue or halt. The interface~$R$ wraps the entire module as a callable API boundary. Goal~$G$ is typically encoded in observations~$O$ or passed as a separate task parameter; for visual simplicity, we show~$O$ as the single input.}
  \label{fig:skill_anatomy}
\end{figure}

This formalization parallels the options framework $(I, \pi, \beta)$ of Sutton et al.~\cite{sutton1999options}, where our $C$ corresponds to the initiation set $I$ and $T$ to the termination condition $\beta$. The interface $R$ builds on the options framework by making skills explicitly invocable, which is necessary for runtime composition. RL options are instead chosen implicitly by a meta-policy, so they do not address this requirement. Fig.\ref{fig:skill_anatomy} shows the resulting four-component architecture.

\subsection{Skills versus Related Abstractions}
\label{sec:def:comparison}

We compares agentic skills with four related concepts (Table~\ref{tab:concept_unification}): unit of reuse, execution semantics, verification surface, composability, and governance surface.

\begin{table*}[t]
\centering
\caption{Concept unification: agentic skills versus related abstractions in LLM agent systems.}
\label{tab:concept_unification}
\resizebox{\textwidth}{!}{%
\begin{tabular}{c|c|c|c|c|c}
\toprule
\textbf{Abstraction} & \textbf{Unit of Reuse} & \textbf{Execution Semantics} & \textbf{Verification Surface} & \textbf{Composability} & \textbf{Governance Surface} \\
\midrule
{Tool} & Single API call & Stateless, single invocation & Input/output schema & Sequential chaining & Permission per tool \\
{Plan} & Task decomposition & One-time reasoning scaffold & Step consistency & Hierarchical decomposition & N/A (ephemeral) \\
{Episodic memory} & Stored observation & Retrieval, no direct execution & Relevance, recency & Indirect (informs reasoning) & Access control on store \\
{Prompt template} & Text fragment & Injected into context window & Output quality & String concatenation & Template authorship \\
\midrule
\textbf{Agentic skill} & Procedural module &  \makecell{ Callable workflow \\with termination} & \makecell{ Outcome  \\correctness, safety} &  \makecell{ Hierarchical, \\DAG, recursive} & \makecell{ Trust tier, sandboxing, \\ provenance} \\
\bottomrule
\end{tabular}%
}
\end{table*}

\para{Tools}
A tool is an atomic primitive (e.g., a web-search API or a file-write function) with a fixed interface and no internal decision-making. Prior work such as Toolformer~\cite{schick2024toolformer} shows that LLMs can learn to invoke tools autonomously, but such behavior typically remains at the level of single calls. A skill may \emph{invoke} tools, but extends them with applicability logic, multi-step sequencing, and explicit termination criteria. Conceptually, the distinction resembles that between a system call and a library routine in software engineering.

\para{Plans} A plan is a reasoning artifact produced by the agent to decompose a task into sub-goals. Plans are typically one-time, session-scoped, and not directly executable without further interpretation. Skills, by contrast, persist across sessions, carry executable policies, and expose callable interfaces. A plan may \emph{select} skills, but a skill is not a plan.

\para{Memory} Episodic and semantic memory systems store observations and facts for later retrieval~\cite{packer2023memgpt,hatalis2023memory,ma2025sok}. Skills are a form of \emph{procedural} memory: they encode \emph{how} to act, not \emph{what} happened. The relationship between declarative memory and procedural skills in LLM agents mirrors the distinction drawn in cognitive psychology between knowing-that and knowing-how~\cite{anderson2004actr}.

\para{Prompt templates} Prompt templates are static text fragments injected into the context window~\cite{white2023prompt}. They lack applicability conditions, termination logic, and callable interfaces. A skill may \emph{contain} a prompt template as part of its policy $\pi$, but a template alone does not constitute a skill.

\para{Classical AI planning formalisms} The skill abstraction also connects to classical AI planning. In Hierarchical Task Networks (HTNs)~\cite{nau2003shop2}, methods decompose tasks into sub-tasks with preconditions, mirroring our hierarchical composition (\S\ref{sec:composition:hierarchical}). BDI (Belief-Desire-Intention) architectures~\cite{rao1995bdi} use reusable plan recipes with context conditions, which align with $C$ and $\pi$. STRIPS/PDDL actions~\cite{fikes1971strips} make preconditions and effects explicit, which anticipates our applicability and termination conditions. The main difference is representational: LLM-based skills act on natural-language observations and can encode policies as NL instructions or hybrid artifacts, while classical formalisms assume symbolic state. We keep the formalization representation-agnostic to connect these lines of work.

\subsection{Skills as Procedural Memory}
\label{sec:def:procmem}

Cognitive science provides a useful lens for understanding why skills matter. Anderson’s ACT-R theory~\cite{anderson2004actr} distinguishes declarative memory (facts and episodes) from procedural memory (production rules that encode condition–action pairs). Experts differ from novices less in what they know than in the richness of their procedural repertoire: action patterns that trigger automatically when conditions are met, freeing working memory for higher-level reasoning.

LLM agents face an analogous challenge. Without a skill layer, every task requires the agent to reason from first principles within a limited context window, consuming tokens to re-derive procedures that could be stored and retrieved. Skills serve as the procedural memory of the agent, compressing learned procedures into reusable modules that reduce the cognitive load on the model's context window, analogous to how chunking in human expertise compresses multi-step procedures into single retrievable units~\cite{chase1973chess}.

This framing has a practical implication: the value of a skill is not merely convenience but \emph{reliability}. A curated skill that has been verified across multiple contexts is more reliable than an ad-hoc plan generated on the fly, for the same reason that a tested library function is more reliable than inline code. Recent empirical evidence supports this intuition: the SkillsBench benchmark~\cite{skillsbench2025} demonstrates that curated skills raise agent pass rates by 16.2~percentage points on average, while self-generated skills degrade performance by 1.3\,pp, encoding incorrect or overly specific heuristics. Notably, a smaller model equipped with curated skills can outperform a larger model operating without them. One interpretation is that procedural memory serves as an efficiency multiplier and partially substitute for model scale.

\begin{figure*}[t]
  \centering
  \includegraphics[width=0.8\linewidth]{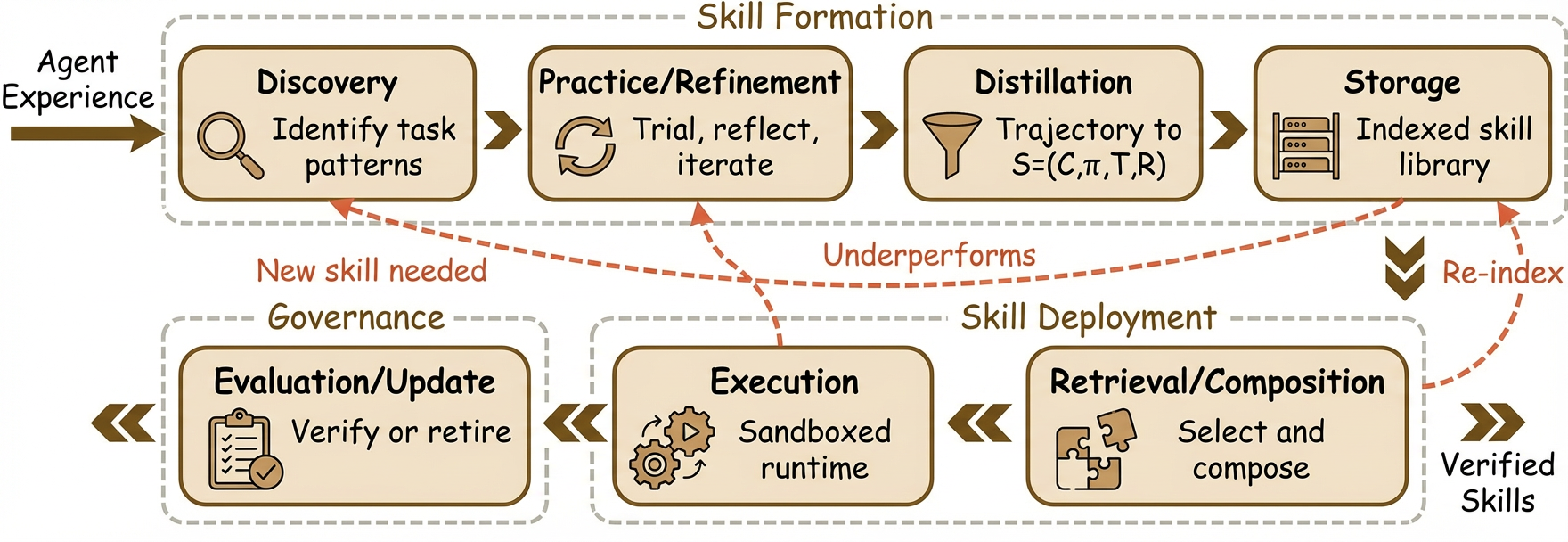}
  \caption{The agentic skill lifecycle. Solid arrows indicate the primary forward path; dashed arrows indicate feedback loops for refinement and retirement. Each stage corresponds to a body of research surveyed in this paper.}
  \label{fig:lifecycle}
\end{figure*}

\section{Methodology}
\label{sec:methodology}

This section describes the systematic process used to collect and analyze the literature on agentic skills in LLM agent systems, and the methodology through which our taxonomies were developed.

\subsection{Literature Search and Selection}
\label{sec:methodology:search}

We ran a structured search across six databases: Google Scholar, Semantic Scholar, DBLP, ACM Digital Library, IEEE Xplore, and arXiv. We used keyword queries (including \emph{agent skills}, \emph{skill learning LLM}, \emph{reusable agent behaviors}, \emph{procedural knowledge agents}, \emph{tool composition LLM}, \emph{agent libraries}, and \emph{hierarchical agent policies}) and followed citations forward and backward from seed papers (Voyager~\cite{wang2023voyager}, ReAct~\cite{yao2023react}, Reflexion~\cite{shinn2023reflexion}, SWE-agent~\cite{yang2024sweagent}).

The search covered publications from January 2020 through February 2025 for LLM agent systems. We include one concurrent work, SkillsBench~\cite{skillsbench2025} (February 2026), as a notable exception given its direct relevance to skill evaluation; all other primary sources fall within the stated window. Foundational works from cognitive science~\cite{anderson2004actr,chase1973chess}, cognitive architectures~\cite{laird2012soar}, and reinforcement learning~\cite{sutton1999options} were included regardless of date to ground the skill abstraction in established theory.

\para{Inclusion criteria} A paper was included if it satisfies at least one of the following: (i)~it introduces, implements, or evaluates reusable procedural capabilities for LLM-based or language-conditioned agents; (ii)~it addresses at least one lifecycle stage (discovery, refinement, distillation, storage, retrieval, execution, or evaluation) of agent procedural knowledge; or (iii)~it provides a benchmark environment in which agent skills can be measured.

\para{Exclusion criteria} We excluded works that focus exclusively on single-turn tool calling without procedural composition, pure prompt engineering without skill persistence or reuse, and multi-agent coordination papers that do not involve a skill abstraction. We also excluded papers focusing solely on fine-tuning for instruction following without an explicit skill representation.

\subsection{Corpus and Analysis}
\label{sec:methodology:corpus}

The initial search yielded approximately 180 candidate papers. After applying the inclusion and exclusion criteria, we retained 65 papers for detailed analysis. Of these, 24 systems are analyzed in depth through the mapping tables (Tables~\ref{tab:lifecycle_mapping} and~\ref{tab:taxonomy_master}), and the remaining papers inform the analysis across lifecycle stages, design patterns, security, and evaluation. The corpus spans eight benchmark environments, seven design patterns, and five representation categories.

\subsection{Taxonomy Development}
\label{sec:methodology:taxonomy}

Both taxonomies were developed through an iterative bottom-up process. We first compiled a feature matrix for each analyzed system, recording its skill representation, acquisition method, execution model, storage mechanism, and governance features. Recurring clusters in this matrix suggested the seven design patterns (\S\ref{sec:patterns}) and the five representation categories (\S\ref{sec:taxonomy:repr}).

We tested each candidate taxonomy against the full corpus, refining categories through three revision cycles until every deeply-analyzed system could be classified without forcing. The representation $\times$ scope taxonomy was developed orthogonally: scope categories emerged from the environments addressed by the analyzed systems. Their orthogonality with representation was validated by confirming that examples exist across most cells of the resulting matrix.

The design patterns are deliberately \emph{non-exclusive}: real systems often combine multiple patterns (e.g., a marketplace-distributed plugin using metadata-driven loading with hybrid NL+code implementation). We treat composability as a feature of the pattern framework rather than a taxonomy deficiency, as mutually exclusive patterns would not reflect how skills are deployed in practice.

\section{Skill Lifecycle Model}
\label{sec:lifecycle}

We organize the literature around a lifecycle model that traces an agentic skill from initial formation to eventual retirement. Rather than viewing skills as static artifacts, this model treats them as evolving system components shaped by interaction, feedback, and deployment constraints. The lifecycle comprises seven stages, depicted in Fig.\ref{fig:lifecycle}:

\begin{itemize}[leftmargin=*,nosep]
  \item \textit{Discovery}: identifying recurring task patterns, failure modes, or workflow bottlenecks that justify encapsulating behavior into a reusable skill.
  \item \textit{Practice/Refinement}: iteratively improving candidate skills through trial-and-error execution, reflection, and external feedback, allowing policies and prompts to stabilize across repeated use.
  \item \textit{Distillation}: extracting a stable and generalizable procedure from trajectories or demonstrations and packaging it into the $(C, \pi, T, R)$ tuple together with descriptive metadata and usage constraints.
  \item \textit{Storage}: persisting the skill within a library or repository, accompanied by indexing, versioning, and metadata that enable efficient retrieval and governance.
  \item \textit{Retrieval/Composition}: selecting relevant skills at runtime and composing them into higher-level workflows, often requiring compatibility checks across interfaces, contexts, and dependencies.
  \item \textit{Execution}: running the skill policy within the agent’s action loop under sandboxing, permission controls, and resource constraints that bound potential side effects.
  \item \textit{Evaluation/Update}: monitoring performance after deployment, detecting drift or failure, and revising, replacing, or retiring skills as environments and requirements evolve.
\end{itemize}

The lifecycle is not strictly linear. Feedback loops connect evaluation back to practice (when a skill underperforms), retrieval back to storage (when indexing fails to surface relevant skills), and execution back to discovery (when runtime failures reveal the need for new skills). Table~\ref{tab:lifecycle_mapping} maps representative systems to lifecycle contributions.

\begin{table*}[t]
\centering
\caption{Lifecycle mapping: representative systems and their primary contributions to skill lifecycle stages. ``\cmark'' = primary focus; ``\pmark'' = partial coverage.}
\label{tab:lifecycle_mapping}
\resizebox{\textwidth}{!}{%
\begin{tabular}{c|c|c|ccccccc|c}
\toprule
\textbf{System/Paper} & \textbf{Environment} & \textbf{Signal} & \rotatebox{70}{\textbf{Discovery}} & \rotatebox{70}{\textbf{Practice}} & \rotatebox{70}{\textbf{Distillation}} & \rotatebox{70}{\textbf{Storage}} & \rotatebox{70}{\textbf{Retrieval}} & \rotatebox{70}{\textbf{Execution}} & \rotatebox{70}{\textbf{Evaluation}} & \textbf{Representation} \\
\midrule
Voyager~\cite{wang2023voyager} & Minecraft & Self-verify & \cmark & \cmark & \cmark & \cmark & \cmark & \cmark & \pmark & Code \\
JARVIS-1~\cite{wang2023jarvis1} & Minecraft & Multimodal & \pmark & \cmark & \pmark & \cmark & \cmark & \cmark & \pmark & Hybrid \\
DEPS~\cite{wang2023deps} & Minecraft & LLM planner & \cmark & \cmark & \pmark & \pmark & \pmark & \cmark & \xmark & NL \\
Reflexion~\cite{shinn2023reflexion} & Multi & Verbal RL & \xmark & \cmark & \pmark & \pmark & \xmark & \cmark & \cmark & NL \\
Skill-it!~\cite{chen2023skillit} & Language & Curriculum & \cmark & \cmark & \cmark & \xmark & \xmark & \pmark & \cmark & Latent \\
SWE-agent~\cite{yang2024sweagent} & SWE & Execution & \xmark & \pmark & \xmark & \xmark & \xmark & \cmark & \cmark & Code \\
AppAgent~\cite{zhang2023appagent} & Mobile & Demonstrations & \cmark & \cmark & \cmark & \cmark & \cmark & \cmark & \pmark & Hybrid \\
WebArena agents~\cite{zhou2024webarena} & Web & Task reward & \xmark & \pmark & \xmark & \xmark & \xmark & \cmark & \cmark & NL/Code \\
CRADLE~\cite{tan2024cradle} & Games & Multi-source & \cmark & \cmark & \cmark & \cmark & \cmark & \cmark & \cmark & Hybrid \\
MemGPT~\cite{packer2023memgpt} & Multi & Self-edit & \xmark & \xmark & \xmark & \cmark & \cmark & \pmark & \xmark & NL \\
AgentTuning~\cite{zeng2023agenttuning} & Multi & SFT & \pmark & \pmark & \cmark & \xmark & \xmark & \cmark & \cmark & Policy \\
CodeAct~\cite{wang2024codeact} & Multi & Code exec & \xmark & \xmark & \xmark & \xmark & \xmark & \cmark & \cmark & Code \\
FireAct~\cite{chen2023fireact} & Multi & Trajectories & \xmark & \pmark & \cmark & \xmark & \xmark & \cmark & \cmark & Policy \\
HuggingGPT~\cite{shen2023hugginggpt} & Multi & API routing & \xmark & \xmark & \xmark & \cmark & \cmark & \cmark & \pmark & Hybrid \\
TaskWeaver~\cite{qin2023taskweaver} & SWE & Code-first & \xmark & \pmark & \pmark & \cmark & \cmark & \cmark & \pmark & Code \\
\midrule
SayCan~\cite{ahn2022saycan} & Robotics & Affordance & \cmark & \pmark & \pmark & \cmark & \cmark & \cmark & \pmark & Hybrid \\
MetaGPT~\cite{hong2024metagpt} & Multi & Role assign & \pmark & \pmark & \cmark & \cmark & \cmark & \cmark & \pmark & Code \\
Generative Agents~\cite{park2023generative} & Social & Self-observe & \pmark & \pmark & \pmark & \cmark & \cmark & \cmark & \pmark & NL \\
Eureka~\cite{ma2024eureka} & Robotics & Reward search & \cmark & \cmark & \cmark & \pmark & \xmark & \cmark & \cmark & Code \\
\bottomrule
\end{tabular}%
}
\end{table*}

\subsection{Discovery}
\label{sec:lifecycle:discovery}

Skill discovery is the process of identifying recurring task patterns that warrant encapsulation into a reusable module. In Voyager~\cite{wang2023voyager}, discovery is driven by a curriculum mechanism that proposes increasingly complex tasks in Minecraft; when the agent succeeds at a novel task, the solution trajectory becomes a candidate skill. DEPS~\cite{wang2023deps} discovers skills through plan decomposition: a high-level planner identifies sub-goals, and repeated sub-goal patterns are promoted to skills. AppAgent~\cite{zhang2023appagent} discovers skills through user demonstrations on mobile interfaces, identifying reusable interaction patterns across applications.

In the robotics domain, SayCan~\cite{ahn2022saycan} discovers executable skills by grounding language instructions in robot affordances: the system scores candidate skills by both language relevance and physical feasibility, effectively discovering which skills apply in a given context. DECKARD~\cite{nottingham2023deckard} uses language-guided world models to discover skills through embodied decision making, imagining plans before executing them.

A key open question is \emph{unsupervised} discovery: identifying skill boundaries without human-provided task definitions or explicit success signals. Current systems rely on either pre-defined task curricula or human demonstrations to seed the discovery process.

\subsection{Practice, Refinement, and Distillation}
\label{sec:lifecycle:practice}

Once a candidate skill is identified, it must be refined into a reliable procedure. Reflexion~\cite{shinn2023reflexion} demonstrates a verbal reinforcement learning loop where the agent reflects on failed attempts and generates textual feedback that guides subsequent trials. This reflection mechanism serves as a practice loop that improves skill reliability without parameter updates.

Distillation converts raw traces into compact, generalizable skill representations. AgentTuning~\cite{zeng2023agenttuning} distills traces from GPT-4 into smaller models through supervised fine-tuning, producing agents with internalized skills. FireAct~\cite{chen2023fireact} fine-tunes agents on diverse ReAct-style traces, distilling multi-step reasoning patterns into model weights.

Inner Monologue~\cite{huang2022innermonologue} extends verbal feedback to embodied agents by using language-based scene descriptions and success signals to iteratively refine robotic action sequences. Eureka~\cite{ma2024eureka} shows that LLMs can autonomously design reward functions for robotic skill acquisition through evolutionary search, achieving human-level performance and effectively automating the practice-and-refine loop for physical skills.

The distinction between practice and distillation is important: practice improves a skill's reliability through iteration, while distillation changes its representation (e.g., from a verbose trajectory to a compact code function or from prompt-based instructions to model weights).

\subsection{Storage and Retrieval}
\label{sec:lifecycle:storage}

Skill storage requires indexing mechanisms that support efficient retrieval. Voyager~\cite{wang2023voyager} maintains a skill library indexed by natural-language descriptions, using embedding similarity to retrieve relevant skills for new tasks. CRADLE~\cite{tan2024cradle} extends this with multi-level memory that stores skills alongside episodic context, enabling retrieval based on both task similarity and environmental state.

The storage-retrieval interface is where skill systems intersect with memory architectures. MemGPT~\cite{packer2023memgpt} provides a hierarchical memory system that could serve as infrastructure for skill libraries, with main memory (context window) and archival storage (external database) supporting different access patterns. Generative Agents~\cite{park2023generative} implement a memory architecture for simulated social agents where behavioral patterns, analogous to social skills, are stored and retrieved based on recency, importance, and relevance, providing a model for how skill libraries might integrate with broader memory systems. The challenge is designing retrieval policies that balance precision (returning the most applicable skill) with recall (not missing relevant skills in novel contexts).

\subsection{Execution and Evaluation}
\label{sec:lifecycle:execution}

Execution is the stage where a skill's policy $\pi$ is enacted within the agent's action loop. The execution model varies substantially by skill representation: natural-language skills are injected into the context window, code skills are executed in sandboxed environments, and policy skills operate through learned parameters. CodeAct~\cite{wang2024codeact} demonstrates that representing agent actions as executable Python code, rather than as tool-calling JSON, improves both the expressiveness and the verifiability of skill execution.

Evaluation assesses whether a skill achieves its intended outcome reliably. Deterministic evaluation harnesses, where the environment itself provides ground-truth verification, are preferable to human grading for scalability. SkillsBench~\cite{skillsbench2025} operationalizes this principle by pairing each of its 86~tasks with a deterministic verifier that checks environment state against expected outcomes, enabling reproducible evaluation across 7{,}308~agent trajectories. 

We discuss evaluation in depth in \S\ref{sec:evaluation}.

\section{Design Patterns and Taxonomy}
\label{sec:patterns}

\begin{figure}[t]
  \centering
  \includegraphics[width=\linewidth]{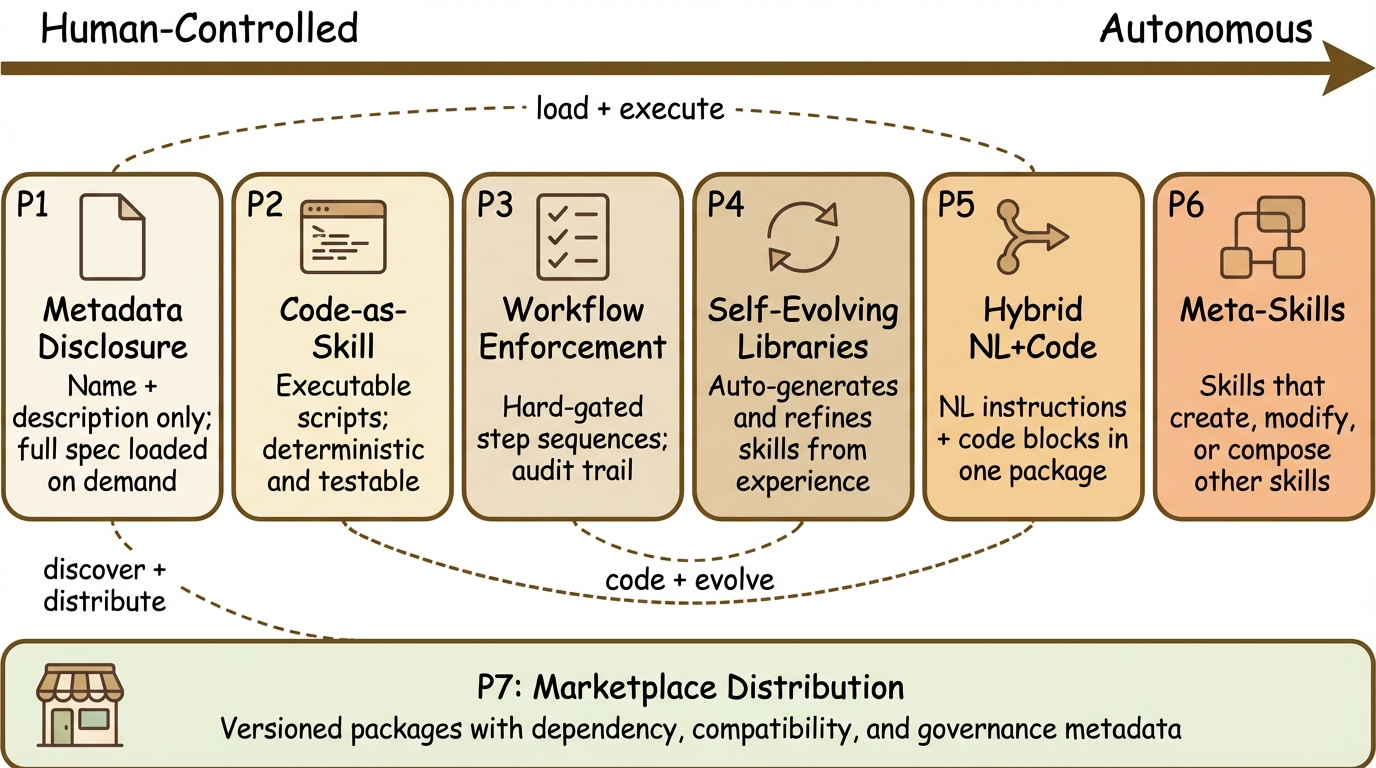}
  \caption{Seven design patterns for agentic skills arranged along an autonomy spectrum, from human-controlled metadata disclosure~(P1) to fully autonomous meta-skills~(P6). Marketplace distribution~(P7) spans the full spectrum as a cross-cutting distribution mechanism. Dashed lines indicate commonly combined patterns.}
  \label{fig:pattern_spectrum}
\end{figure}

We classify the emerging skill landscape along two complementary dimensions. First, we identify seven \emph{design patterns} that describe how skills are packaged, loaded, and executed at the system level. Second, we develop a \emph{representation $\times$ scope taxonomy} (\S\ref{sec:taxonomy}) that describes what skills are and where they operate. Fig.\ref{fig:pattern_spectrum} arranges the seven patterns along an autonomy axis. This section presents both, beginning with the design patterns summarized in Table~\ref{tab:patterns}.

\begin{table*}[t]
\centering
\caption{Seven design patterns for agentic skills, with representative systems, strengths, and governance considerations.}
\label{tab:patterns}
\resizebox{\textwidth}{!}{%
\begin{tabular}{c|L{2.8cm}|L{3.32cm}|c|L{2.9cm}|L{2.7cm}|L{2.5cm}}
\toprule
\textbf{\#} & \textbf{Pattern} & \textbf{Representative Systems} & \textbf{Representation} & \textbf{Strength} & \textbf{Weakness} & \textbf{Primary Risk} \\
\midrule
\rowcolor{pone!50}
1 & Metadata-driven progressive disclosure & Claude Code, Semantic Kernel, LangChain & NL + metadata & Token efficiency; scales to large libraries & Retrieval quality depends on metadata & Metadata poisoning \\
2 & Code-as-skill (executable scripts) & Voyager, CodeAct, SWE-agent & Code & Deterministic; testable; composable & Requires sandbox; brittle to API changes & Code injection \\
\rowcolor{pone!50}
3 & Workflow enforcement & TDD agents, LATS, systematic debuggers & NL + rules & Reliability through gating; auditable & Rigid; may over-constrain agent & Rule bypass via prompt injection \\
4 & Self-evolving skill libraries & Voyager, DEPS, CRADLE & Code + NL & Adapts to new tasks; improves with use & Quality control of self-generated skills & Skill drift; poisoned distillation \\
\rowcolor{pone!50}
5 & Hybrid NL+code macros & Claude skills, ReAct prompts & NL + code + refs & Flexible; human readable yet executable & Ambiguity at NL/code boundary & Inconsistent interpretation \\
6 & Meta-skills (skills that create skills) & Self-Instruct, skill generators & NL / hybrid & Scales skill library; reduces human effort & Bootstrapping quality ceiling & Recursive error amplification \\
\rowcolor{pone!50}
7 & Plugin / marketplace distribution & OpenAI GPT Store, MCP servers, ClawHub, npm/pip & Any (packaged) & Ecosystem growth; community contribution & Supply-chain trust; version compat & Malicious packages (cf.\ ClawHavoc) \\
\bottomrule
\end{tabular}%
}
\end{table*}

\subsection{Why Design Patterns?}
\label{sec:patterns:why}

Software engineering has long treated recurring design patterns as worth documenting for both practice and research~\cite{gamma1994designpatterns}. We follow that approach for agentic skills. A design pattern describes a solution shape that shows up across systems. Here, patterns are \emph{system}-level: they describe how infrastructure manages skills. By contrast, the representation/scope taxonomy is \emph{skill}-level. They complement each other: one pattern (e.g., marketplace distribution) can host skills with any representation and scope.

\subsection{Pattern-1: Metadata-Driven Disclosure}
\label{sec:patterns:metadata}

In Pattern-1, skills are discovered through compact metadata summaries (name, description, trigger conditions) that occupy minimal context. The full instructions are loaded into the agent's context window only when the skill is selected for execution. This two-phase loading strategy addresses a fundamental constraint of LLM agents: the finite context window cannot hold all available skills simultaneously.

Claude Code's skill system exemplifies this pattern. Each skill is registered with a short description and a set of trigger phrases. When the agent determines that a skill is relevant to the current task, it loads the full skill specification, which may include multi-page instructions, reference documents, and execution scripts. The Semantic Kernel framework~\cite{semantickernel2023} implements a similar approach with its plugin discovery mechanism, where function metadata is registered and the full function implementation is invoked only on selection.

The main benefit is scale: an agent can know about hundreds of skills while spending context tokens only on the few it activates. The main risk is metadata quality. If descriptions are wrong or incomplete, retrieval can pick the wrong skill or miss a relevant one.

\subsection{Pattern-2: Code-as-Skill (Executable Scripts)}
\label{sec:patterns:code}

Code-as-skill represents skills as executable programs (Python functions, shell scripts, or domain-specific language programs) that the agent invokes through a runtime interface. Voyager~\cite{wang2023voyager} generates JavaScript functions as skills for Minecraft, stores them in a library, and retrieves them by natural-language description. CodeAct~\cite{wang2024codeact} demonstrates that framing agent actions as executable Python code, rather than as structured JSON tool calls, enables more expressive and verifiable behavior. In robotics, Code as Policies~\cite{liang2023code} generates Python programs for robotic control, and ProgPrompt~\cite{singh2023progprompt} creates situated task plans as executable programs, both treating generated code as reusable skills for physical manipulation.

The key advantage of code skills is \emph{determinism}: given the same inputs, a code skill produces the same outputs, enabling traditional software testing and verification. Code skills also composed of function calls, imports, and control flow. The limitation is brittleness: code skills break when underlying APIs, UI elements, or environmental conditions change, necessitating maintenance and version management (\S\ref{sec:lifecycle:execution}).

\subsection{Pattern-3: Workflow Enforcement}
\label{sec:patterns:workflow}

Workflow-enforcement skills impose hard-gated processes on agent behavior, ensuring that the agent follows a prescribed methodology rather than improvising. A test-driven development (TDD) skill, for example, mandates that the agent write tests before implementation, run the test suite, and iterate until all tests pass. The agent cannot skip or reorder these steps.

LATS (Language Agent Tree Search)~\cite{zhou2023lats} enforces a tree-search workflow that combines planning, acting, and reflection in a structured loop. Systematic debugging skills enforce a diagnosis-before-fix methodology, requiring the agent to reproduce the bug, identify root causes, and verify the fix before declaring success.

This pattern sacrifices flexibility for reliability. By constraining the agent's action space to a proven sequence, workflow enforcement reduces the probability of hallucination-driven shortcuts and provides a clear audit trail. We note that Pattern-3 operates at the controller level: it prescribes \emph{how} the agent executes rather than constituting a reusable skill artifact itself. LATS exemplifies a workflow controller that can host skills from other patterns. The governance surface is the rule set itself: if an attacker can modify the workflow rules (e.g., through prompt injection), the enforcement mechanism is compromised.

\subsection{Pattern-4: Self-Evolving Skill Libraries}
\label{sec:patterns:selfevolving}

Self-evolving skill libraries combine skill execution with automated quality assessment and library maintenance. After each task, the system evaluates whether the agent's behavior produced a successful trajectory worthy of distillation into a new skill or refinement of an existing one.

Voyager~\cite{wang2023voyager} provides a canonical example: it generates code-based skills, validates them through in-game execution, and incorporates verified skills into a persistent library indexed by natural-language descriptions. CRADLE~\cite{tan2024cradle} extends this paradigm with explicit memory management, linking skills to episodic context to enable retrieval based on environmental similarity.

The central tension in self-evolving libraries is quality control. The SkillsBench benchmark~\cite{skillsbench2025} reports that self-generated skills average $-$1.3\,pp relative to skill-free baselines, with only one of five tested configurations showing any improvement, indicating that zero-shot self-generation without iterative verification can degrade performance in open-ended settings. This contrasts with Voyager and Eureka, where self-generated skills succeed in constrained environments with deterministic execution verification, suggesting the viability of self-generation depends critically on domain specificity and the availability of automated verification. Without human oversight or robust verification, self-evolving libraries risk accumulating ``skill debt'' analogous to technical debt in software systems.

\subsection{Pattern-5: Hybrid NL+Code Macros}
\label{sec:patterns:hybrid}

Hybrid skills combine natural-language specifications with executable components within a single package. The natural-language component describes the skill’s purpose, applicability conditions, and high-level logic in human-readable form, while the executable component provides code snippets, reference documents, or tool-calling sequences that implement concrete steps.

This pattern appears in production agent systems where skills must be both human-auditable and machine-executable. Claude Code's skill system, for example, defines skills as markdown documents that include natural-language instructions, code blocks, and references to external assets. The ReAct paradigm~\cite{yao2023react} represents a lightweight version: the agent alternates between natural-language reasoning (``I need to search for X'') and executable actions (search API call), with the interleaving serving as an implicit hybrid skill.

The advantage of hybrid skills is flexibility: the natural-language component provides context and handles edge cases through reasoning, while the code component provides determinism for well-understood steps. The risk is \emph{boundary ambiguity}: when instructions conflict with code, the agent must decide which to follow, creating potential for inconsistent behavior.

\subsection{Pattern-6: Meta-Skills}
\label{sec:patterns:meta}

Meta-skills are skills whose purpose is to create, modify, or compose other skills. A meta-skill might analyze an agent's task history to identify recurring patterns, generate candidate skills from those patterns, and test them against held-out tasks. Self-Instruct~\cite{wang2023selfinstruct} can be viewed through this lens: the LLM generates new instruction-following examples that serve as training data for skill acquisition. CREATOR~\cite{qian2023creator} takes this further by enabling LLMs to create new tools (i.e., code functions) on demand, disentangling abstract reasoning from concrete tool implementation. Eureka~\cite{ma2024eureka} generates reward functions that serve as parameterizations for robotic skills, effectively creating skill specifications through code.

We cite Self-Instruct and CREATOR as precursors: they show the training-time idea of meta-skills, but they are mostly offline. We reserve Pattern-6 in the strict sense for methods that act as runtime-callable generators. Discovery (\S\ref{sec:lifecycle:discovery}) makes the procedural gap explicit; meta-skills are the generative mechanism that fills it. These sit at different levels (lifecycle stage vs.\ design pattern): a meta-skill automates what would otherwise be manual discovery.

Meta-skills let a small seed set of skills grow into a broad library without requiring a matching amount of human work. The risk is \emph{recursive error amplification}: if the meta-skill produces a flawed skill that is subsequently used as input for further skill generation, errors compound. Quality gates at each generation step are essential (\S\ref{sec:security}).

\subsection{Pattern-7: Plugin/Marketplace Distribution}
\label{sec:patterns:marketplace}

The marketplace pattern treats skills as versioned, distributable packages with explicit dependency, compatibility, and governance metadata. The OpenAI GPT Store distributes custom GPT configurations that function as packaged skills. Anthropic's Model Context Protocol (MCP)~\cite{anthropic2024mcp} defines a standardized interface for tool and skill servers, enabling third-party skill distribution with authentication and permission boundaries. ToolLLM~\cite{qin2024toolllm} demonstrates integration with over 16,000 real-world APIs, illustrating the scale that marketplace-style distribution can achieve.

The most striking example of marketplace-scale skill distribution is OpenClaw\openclaw~\cite{openclaw2026}, a viral agent framework built on a four-tool core (read, write, edit, bash) that treats skills as the primary extensibility mechanism. OpenClaw's community skill registry, ClawHub, grew from zero to over 10{,}700 published skills within weeks of launch, while the project itself surpassed 200{,}000 GitHub stars faster than any software repository in history~\cite{clawhavoc2026}. OpenClaw's design philosophy is particularly relevant to our taxonomy: it embraces \emph{self-generated skills} (Pattern-4 $+$ Pattern-6) by encouraging agents to extend themselves through code rather than downloading pre-built extensions. When combined with community distribution (Pattern-7), this creates a dual-source skill library: human-authored community skills alongside agent-authored local skills, both executable with full system access.

In the software ecosystem, analogous patterns include npm packages for JavaScript, pip packages for Python, and plugin systems in IDEs. The marketplace pattern enables community-driven skill creation at scale but introduces \emph{supply-chain risk}: a malicious or compromised skill package can execute arbitrary actions within the agent's permission scope. OpenClaw's explosive growth and the severity of its subsequent security incidents (\S\ref{sec:security:casestudy}) provide a stark illustration of this risk. We analyze these risks in detail in \S\ref{sec:security}.

\subsection{Pattern Trade-offs}
\label{sec:patterns:tradeoffs}

Our patterns represent different points in a multi-dimensional trade-off space. Table~\ref{tab:pattern_tradeoffs} summarizes the key dimensions: \textit{Context cost} measures how many tokens a pattern consumes during active use. \textit{Determinism} reflects the predictability of execution outcomes. \textit{Composability} captures how easily skills following this pattern can be combined into larger workflows. \textit{Governance} surface indicates how amenable the pattern is to auditing, permission control, and provenance tracking.

\begin{table}[t]
\centering
\caption{Pattern trade-off summary across four dimensions. H = High, M = Medium, L = Low.}
\label{tab:pattern_tradeoffs}
\resizebox{\columnwidth}{!}{%
\begin{tabular}{l|c|c|c|c}
\toprule
\multicolumn{1}{c|}{\textbf{Pattern}} & \textbf{Context cost} & \textbf{Determinism} & \textbf{Composability} & \textbf{Governance} \\
\midrule
1: Metadata & L & L & M & M \\
2: Code-as-skill & L & H & H & H \\
3: Workflow & M & H & M & H \\
4: Self-evolving & M & M & M & L \\
5: Hybrid macro & M & M & M & M \\
6: Meta-skill & H & L & L & L \\
7: Marketplace & L & varies & H & M--H \\
\bottomrule
\end{tabular}%
}
\end{table}

\para{Pattern co-occurrence} Systems in our corpus use a median of 2~patterns (range: 1--4). The most common combination is Patterns~1+7 (metadata + marketplace), appearing in 4~systems (HuggingGPT, MetaGPT, AutoGen, ToolLLM). Two systems (Claude Code and OpenClaw) use 4~patterns, representing outliers. Five systems use a single pattern, while twelve use exactly two. The modest co-occurrence rates suggest the patterns capture meaningfully distinct architectural choices rather than collapsing into a single cluster.

No single pattern dominates. Production systems typically combine patterns: a marketplace-distributed plugin (Pattern-7) might use metadata-driven loading (Pattern-1) with hybrid NL+code implementation (Pattern-5) and workflow enforcement for critical steps (Pattern-3).

\para{Computational overhead} The skill layer imposes overhead: retrieval adds latency, instruction loading consumes context tokens, and multi-level composition multiplies both. Table~\ref{tab:pattern_tradeoffs} captures this abstractly as ``context cost,'' but quantifying the latency-accuracy tradeoff of skill-based versus skill-free agents across deployment scenarios remains an open empirical question.

\subsection{Representation $\times$ Scope Taxonomy}
\label{sec:taxonomy}

Complementing the system-level design patterns, we propose an intrinsic taxonomy along two orthogonal axes: \emph{representation} (how the skill's policy is encoded) and \emph{scope} (what environment or task domain the skill operates over). While patterns describe how infrastructure manages skills, this taxonomy operates at the \emph{skill} level.

\subsubsection{\underline{Representation Axis}}
\label{sec:taxonomy:repr}

We identify five representation categories, ordered roughly by increasing formality:

\para{Natural-language skills} The policy $\pi$ is expressed entirely in natural language: step-by-step instructions, standard operating procedures (SOPs), or playbook entries. The agent interprets these instructions through its language understanding capabilities. Natural-language skills are easy to author and audit but are subject to interpretation ambiguity and cannot be verified through traditional testing.

\para{Code-as-skill} The policy $\pi$ is an executable program: a Python function, a shell script, a domain-specific language program, or a Jupyter notebook cell. Code skills offer determinism and testability but require execution infrastructure and are brittle to environmental changes.

\para{Tool macros} A skill defined as a structured sequence of tool calls with parameterization logic. Tool macros occupy a middle ground between natural language (interpreted) and code (executed): they are more constrained than free-form code but more expressive than single tool calls.

\para{Policy-based skills} The policy $\pi$ is a learned parameterized function, i.e., a neural network fine-tuned on trajectories. Policy skills are opaque (hard to inspect/audit) but capture subtle behavioral patterns that resist explicit codification.

\para{Hybrid representations} A skill that combines two or more of the above. For example, a hybrid skill might use natural-language instructions for high-level logic, code blocks for deterministic steps, and an embedding-based retrieval mechanism for contextual adaptation.

\subsubsection{\underline{Scope Axis}}
\label{sec:taxonomy:scope}

We identify six scope categories based on the environment and task domain:

\para{Single-tool skills} Skills that orchestrate a single tool with sophisticated parameterization, error handling, and retry logic. These are the simplest scope but can still exhibit non-trivial procedural complexity (e.g., a database query skill that handles schema variation).

\para{Multi-tool orchestration} Skills that coordinate multiple tools in sequence or parallel to accomplish a composite task (e.g., search $\rightarrow$ extract $\rightarrow$ summarize $\rightarrow$ store).

\para{Web interaction} Skills for navigating web interfaces, filling forms, extracting information from web pages, and completing web-based workflows. Benchmarked by WebArena~\cite{zhou2024webarena} and Mind2Web~\cite{deng2023mind2web}. A challenge unique to web skills is \emph{UI fragility}: interfaces change frequently, breaking skills that depend on specific element selectors or page layouts. Skills encoding high-level intent (``fill in the departure field'') are more resilient than those encoding low-level actions (``click the element with id=departure-input'').

\para{OS/desktop workflows} Skills that operate across multiple desktop applications, managing windows, files, and system settings. Benchmarked by OSWorld~\cite{xie2024osworld}.

\para{Software engineering} Skills for code understanding, bug localization, patch generation, testing, and deployment. Benchmarked by SWE-bench~\cite{jimenez2024swebench}.

\para{Robotics/physical} Skills for controlling physical actuators, navigating physical spaces, and manipulating objects. While this SoK focuses primarily on digital agents, robotics skill libraries provide instructive parallels, particularly for hierarchical skill composition~\cite{ravichandar2020robotlfd}. Recent work demonstrates diverse LLM-driven robotic skills: SayCan~\cite{ahn2022saycan} grounds language instructions in affordance functions to select feasible skills, Code as Policies~\cite{liang2023code} generates executable robot programs from language, ProgPrompt~\cite{singh2023progprompt} creates situated task plans as programs, and Inner Monologue~\cite{huang2022innermonologue} uses language feedback to refine robotic actions iteratively.

\para{Scope and skill value} The scope axis interacts with skill utility in a non-obvious way. SkillsBench~\cite{skillsbench2025} reports that skills yield the largest improvements in healthcare (+51.9\,pp) and manufacturing (+41.9\,pp) but only +4.5\,pp in software engineering and +6.0\,pp in mathematics. This suggests that skills provide the most value in domains where the base model's pretraining data is sparse or insufficiently procedural, while domains with abundant code and mathematical reasoning data in pretraining benefit less from external procedural knowledge.

\subsubsection{\underline{Mapping: Patterns $\times$ Representation $\times$ Scope}}
\label{sec:taxonomy:mapping}

Table~\ref{tab:taxonomy_master} maps representative systems to all three classification dimensions. The mapping reveals that most systems occupy a sparse region of the full space: code-as-skill representation with SWE or web scope using the self-evolving library pattern. Large regions remain unexplored, particularly policy-based skills with marketplace distribution and natural-language skills with workflow enforcement.

\begin{table*}[t]
\centering
\caption{Taxonomy master table: representative systems mapped to design pattern, representation, scope, and key characteristics.}
\label{tab:taxonomy_master}
\resizebox{\textwidth}{!}{%
\begin{tabular}{c|c|c|c|c|c|c|c}
\toprule
\textbf{System} & \textbf{Pattern(s)} & \textbf{Representation} & \textbf{Scope} & \textbf{Acquisition} & \textbf{Execution} & \textbf{Evaluation} & \textbf{Governance} \\
\midrule
Voyager~\cite{wang2023voyager} & 2, 4 & Code & Game/Robotics & Self-practice & Sandbox & Self-verify & None \\
SWE-agent~\cite{yang2024sweagent} & 2, 3 & Code & SWE & Pre-defined & Shell exec & SWE-bench & Sandboxed \\
CodeAct~\cite{wang2024codeact} & 2 & Code & Multi & Pre-defined & Python exec & AgentBench & Sandboxed \\
CRADLE~\cite{tan2024cradle} & 4, 5 & Hybrid & Game & Self-evolving & Multi-source & Task reward & None \\
AppAgent~\cite{zhang2023appagent} & 1, 5 & Hybrid & Mobile & Demonstrations & UI actions & Task success & None \\
WebArena agents~\cite{zhou2024webarena} & 2, 3 & Code/NL & Web & Pre-defined & Browser & Task reward & Sandboxed \\
HuggingGPT~\cite{shen2023hugginggpt} & 1, 7 & Hybrid & Multi & Pre-registered & API routing & Task output & API auth \\
TaskWeaver~\cite{qin2023taskweaver} & 2, 5 & Code & SWE & Human + gen & Python exec & Output verify & Plugin sys \\
Claude Code\footnote{A production agent system; name retained as it is publicly documented.} & 1, 3, 5, 7 & Hybrid & SWE/Multi & Human-authored & Multi-mode & User verify & Trust tiers \\
MemGPT~\cite{packer2023memgpt} & 1 & NL & Multi & Human-defined & Context mgmt & N/A & Access ctrl \\
AgentTuning~\cite{zeng2023agenttuning} & 4 & Policy & Multi & Distillation & Fine-tuned & AgentBench & None \\
LATS~\cite{zhou2023lats} & 3 & NL + rules & Multi & Pre-defined & Tree search & Task reward & None \\
Reflexion~\cite{shinn2023reflexion} & 3\textsuperscript{\dag} & NL & Multi & Self-practice & Verbal RL & Task reward & None \\
\midrule
MetaGPT~\cite{hong2024metagpt} & 1, 7 & Code & Multi & Role generation & Multi-agent & Task output & Role-based \\
SayCan~\cite{ahn2022saycan} & 1, 2 & Hybrid & Robotics & Affordance grounding & Grounded exec & Task success & None \\
AutoGen~\cite{wu2024autogen} & 1, 7 & Hybrid & Multi & Pre-defined + gen & Multi-agent & Conversation & Protocol \\
Generative Agents~\cite{park2023generative} & 1, 4\textsuperscript{\ddag} & NL & Social & Self-observed & Memory retrieval & Behavioral & None \\
ToolLLM~\cite{qin2024toolllm} & 1, 7 & Hybrid & Multi & API crawling & API routing & ToolEval & API auth \\
OpenClaw~\cite{openclaw2026} & 2, 4, 6, 7 & Code/Hybrid & Multi & Self-generated + community & Bash/code exec & User verify & ClawHub + VirusTotal \\
\bottomrule
\multicolumn{8}{l}{\textsuperscript{\dag}\footnotesize Reflexion performs transient in-context refinement without persistent library updates; we classify it under Pattern-3 only.} \\
\multicolumn{8}{l}{\textsuperscript{\ddag}\footnotesize Generative Agents' behavioral patterns are closer to episodic memory than to skills as formally defined; included as an illustrative boundary case.} \\
\end{tabular}%
}
\end{table*}

\section{Acquisition, Composition, Orchestration}
\label{sec:acquisition}

We address two complementary questions: how agents \emph{acquire} skills, and how they \emph{compose} and \emph{orchestrate} acquired skills at runtime. We begin with five acquisition modes, ordered from most to least human involvement.

\subsection{Human-Authored Skills}
\label{sec:acquisition:human}

The simplest acquisition mode is human authorship. A domain expert writes a skill specification (e.g., a standard operating procedure, a code function, or a hybrid document) and registers it in the agent's skill library. Many production systems (e.g., Claude Code and enterprise automation platforms) rely on human-authored skills because they are easier to validate and assign accountability for.

Human authorship scales poorly but produces high-reliability skills. The trade-off is explicit: each skill requires human labor to create, test, and maintain, but the resulting skills are grounded in domain expertise and can be audited before deployment.

\subsection{Demonstration Distillation}
\label{sec:acquisition:distillation}

Demonstration distillation extracts reusable procedures from observed trajectories. The input may be human demonstrations~\cite{zhang2023appagent}, expert agent traces~\cite{zeng2023agenttuning}, or successful task completions from the agent itself~\cite{wang2023voyager}. The key challenge is \emph{generalization}: a trajectory that solved one specific instance must be abstracted into a skill that handles the broader class.

AgentTuning~\cite{zeng2023agenttuning} collects interaction trajectories from GPT-4 across diverse agent tasks and uses them to fine-tune Llama models, effectively distilling procedural knowledge into model weights. FireAct~\cite{chen2023fireact} fine-tunes language models on ReAct-style trajectories, distilling the reasoning-acting pattern into an internalized skill.

\subsection{Self-Practice and Exploration}
\label{sec:acquisition:selfpractice}

Self-practice acquisition allows the agent to discover and refine skills through autonomous interaction with the environment. Voyager~\cite{wang2023voyager} implements this through a curriculum-driven exploration loop: the agent proposes tasks, attempts them, evaluates success, and stores verified solutions as skills.

Reflexion~\cite{shinn2023reflexion} refines agent behavior through verbal self-reflection: after a failed attempt, the agent generates a textual analysis of what went wrong and uses this analysis to guide the next attempt. While Reflexion does not explicitly produce persistent skills, the reflection mechanism can be viewed as transient skill refinement within an episode.

AutoGPT~\cite{autogpt2023} popularized the paradigm of fully autonomous agents that set their own sub-goals and practice iteratively, though without explicit skill persistence across sessions. DECKARD~\cite{nottingham2023deckard} combines language-guided world models with embodied exploration, imagining and evaluating plans before executing them in game environments.

The self-practice mode enables continual learning~\cite{wang2024comprehensive} without human supervision but introduces quality risk. Without external verification, agents may converge on locally optimal but globally suboptimal procedures, or worse, on procedures that succeed through exploitation of environment quirks rather than genuine task completion.

\subsection{Curriculum and Feedback}
\label{sec:acquisition:curriculum}

Curriculum-based acquisition structures the skill learning process through progressively harder tasks. Skill-it!~\cite{chen2023skillit} provides a theoretical framework for curriculum design in skill learning, demonstrating that training on an ordered sequence of skills improves sample efficiency compared to random ordering.

Feedback signals can come from humans (corrections, preferences), AI judges (LLM-based evaluators), or reward models trained on human preferences. The choice of feedback signal affects both the quality and the scalability of skill acquisition: human feedback is high-quality but expensive, AI judges are scalable but may miss subtle errors, and reward models generalize from limited human data but can be exploited through reward hacking.

\subsection{Meta-Skills and Self-Evolving Libraries}
\label{sec:acquisition:meta}

The most autonomous acquisition mode uses meta-skills (Pattern-6) to generate new skills from existing ones. A meta-skill might analyze an agent's failure cases, identify missing capabilities, and generate candidate skills to fill those gaps. Self-Instruct~\cite{wang2023selfinstruct} demonstrates a related approach: using an LLM to generate new instruction-following examples from a seed set, effectively bootstrapping a skill library from a small initial collection. CREATOR~\cite{qian2023creator} enables LLMs to create new tools on demand, and Eureka~\cite{ma2024eureka} generates reward functions that parameterize robotic skills, both exemplifying meta-skill acquisition at different levels of abstraction.

Self-evolving libraries combine meta-skill generation with automated quality assessment, creating a closed loop in which the skill library grows and improves without human intervention. The primary risk is the \emph{quality ceiling problem}: without external grounding, the library cannot exceed the capability of the meta-skill itself, and errors in early generations may propagate through subsequent ones.

\subsection{Skill composition and orchestration.}\label{sec:composition}
\begin{figure}[t]
  \centering
  \includegraphics[width=0.5\columnwidth]{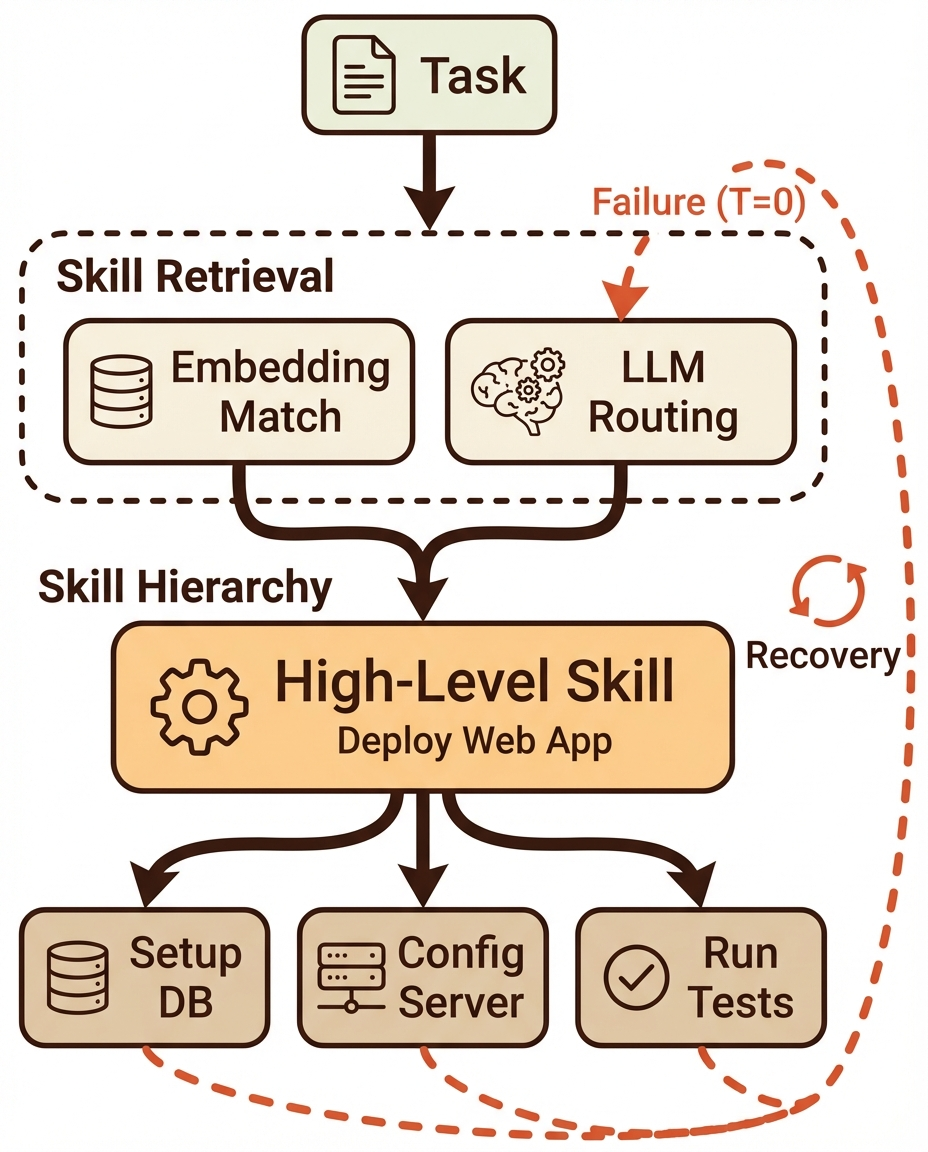}
  \caption{Skill composition and orchestration. Tasks are matched to skills via embedding-based retrieval or LLM-mediated routing. Selected skills decompose hierarchically into sub-skills. Dashed arrows indicate failure recovery paths that trigger re-retrieval or alternative skill selection.}
  \label{fig:skill_composition}
\end{figure}

Individual skills rarely suffice for complex tasks. Fig.\ref{fig:skill_composition} illustrates the composition architecture. The remainder of this section addresses how skills are combined, routed, and managed during multi-step execution.

\subsubsection{Hierarchical Skill Structures}
\label{sec:composition:hierarchical}

Skills organize into hierarchies: a high-level skill (e.g., ``deploy a web application'') invokes mid-level skills (``set up database,'' ``configure server,'' ``run tests''), which in turn invoke low-level skills (``execute SQL migration,'' ``write Nginx config''). This hierarchical structure mirrors the option framework in reinforcement learning~\cite{sutton1999options}, where temporally extended actions (options) compose atomic actions into reusable behavioral modules.

In the LLM agent context, hierarchical composition is typically managed through a planning layer that decomposes tasks and routes sub-tasks to appropriate skills. HuggingGPT~\cite{shen2023hugginggpt} demonstrates this at the tool level, using an LLM planner to decompose requests into sub-tasks routed to specialized Hugging Face models. The same architecture applies to skills: a planner selects and sequences skills based on task requirements and skill metadata.

\para{Runtime skill selection and routing}\label{sec:composition:routing}
When multiple skills could apply to a given context, the agent must select the most appropriate one. Two routing strategies dominate:
\begin{itemize}
    \item \textit{Embedding-based retrieval} The task description is embedded and compared against skill description embeddings. The top-$k$ matching skills are loaded into the context window for the agent to evaluate. Voyager~\cite{wang2023voyager} and AppAgent~\cite{zhang2023appagent} use this approach.

    \item \textit{LLM-mediated routing} The agent itself reasons about which skill to invoke, based on skill metadata loaded through progressive disclosure (Pattern-1). This approach is more flexible than embedding retrieval but consumes additional inference tokens and is subject to the agent's reasoning quality.

\end{itemize}

Hybrid strategies combine both: embedding retrieval narrows the candidate set, and the agent's reasoning selects the final skill. This two-stage approach balances recall (embedding search surfaces relevant candidates) with precision (LLM reasoning evaluates fit).

\textit{Skill conflict resolution.}
When multiple skills are simultaneously applicable ($C_1(o,g)=1$ and $C_2(o,g)=1$), the agent requires a tie-breaking mechanism. Current systems typically rely on ranking heuristics such as embedding similarity or ad hoc LLM judgment, but they lack an explicit conflict-resolution policy. A principled approach, analogous to method specificity in HTNs~\cite{nau2003shop2} or rule priority in production systems, remains an open research problem.

\para{Failure recovery}\label{sec:composition:failure} Failure recovery in skill-based agents can itself be modeled as a skill. When the termination condition $T$ signals failure, a recovery skill is invoked to diagnose the cause and decide whether to retry, backtrack to a prior state, or escalate to an alternative strategy.

LATS~\cite{zhou2023lats} implements recovery through tree search: when a branch fails, the system backtracks and explores alternative action sequences. Reflexion~\cite{shinn2023reflexion} uses verbal reflection as a recovery mechanism, generating natural-language analysis of failures that guides subsequent attempts.

Treating recovery as a first-class skill has governance implications: the recovery skill must be at least as trusted as the skill it is recovering, since it operates in the same execution context and may need to undo or compensate for the failed skill's actions.

\para{Multi-agent skill sharing}\label{sec:composition:multiagent} In multi-agent systems, skills can be shared across agents through common skill repositories. MetaGPT~\cite{hong2024metagpt} assigns specialized roles (product manager, architect, engineer) to different agents, each equipped with role-specific skills that compose into a software development workflow. AutoGen~\cite{wu2024autogen} enables multi-agent conversations where agents with different skill profiles collaborate through structured dialogue protocols. ProAgent~\cite{zhang2024proagent} builds proactive cooperative agents that anticipate teammates' actions and adapt their skill execution accordingly. This enables division of labor: different agents specialize in different skill sets, and tasks are routed to the agent with the most relevant skills. However, shared skill repositories introduce cross-agent security concerns (\S\ref{sec:security}): a compromised skill in a shared repository affects all agents that consume it.

\section{Security, Trust, and Governance of Skills}
\label{sec:security}

The skill layer introduces a new attack surface for LLM agents~\cite{shahriar2025survey}. Skills are code or instructions that influence agent behavior; a compromised skill can steer an agent toward malicious outcomes while appearing benign at the metadata level. This section systematizes threats, mitigations, and governance mechanisms specific to the skill layer.

\subsection{Threat Model}
\label{sec:security:threats}

We identify six primary threat categories:

\para{Poisoned skill retrieval} An attacker crafts skill metadata to cause the retrieval mechanism to surface a malicious skill in response to benign queries. This is analogous to SEO poisoning in web search. The attack exploits Pattern-1 (metadata-driven disclosure): if the retrieval mechanism relies solely on embedding similarity, adversarial metadata can manipulate ranking.

\para{Malicious skill payloads} A skill's policy $\pi$ contains instructions or code that perform unauthorized actions when executed. In code skills (Pattern-2), this resembles supply-chain attacks in traditional software~\cite{ladisa2023supplychain}. In natural-language skills (Pattern-5), the payload is a form of prompt injection: instructions embedded within the skill text that redirect agent behavior.

\para{Cross-tenant leakage} In multi-agent or multi-user systems with shared skill repositories, skills authored by one tenant may access data or resources belonging to another. This risk is acute in enterprise deployments where multiple teams share agent infrastructure: a skill authored by Team~A should not inadvertently access Team~B's data, requiring per-tenant sandboxing with permission boundaries enforced by the execution runtime rather than by the skill itself.

\para{Skill drift exploitation} Skills that were safe at authoring time may become unsafe as environment evolves. An attacker who controls part of the environment (e.g., a web page that a skill navigates) can manipulate environments to change the skill's behavior without modifying the skill itself.

\para{Confused deputy via environmental injection}
An agent processing untrusted observations (e.g., web pages or user documents) may encounter adversarial instructions that coerce it into misusing an otherwise benign, privileged skill. The skill itself remains uncompromised; instead, the attack exploits the data–control boundary between the observation space $O$ and skill invocation. This vector differs from malicious skill payloads, where the attack resides within the skill itself, and is particularly dangerous because it bypasses skill-level trust verification entirely.

\para{Applicability condition poisoning} An attacker manipulates the input to $C$ such that a malicious or inappropriate skill returns $C(o,g)=1$ universally, activating in contexts where it should not. This can occur through metadata poisoning (Pattern-1) or through adversarial environmental states that trigger overbroad applicability predicates. The formal model's reliance on $C$ for skill selection makes this a direct attack on the skill abstraction itself.

\subsection{Trust Tiers and Progressive Disclosure}
\label{sec:security:trust}

\begin{figure}[t]
  \centering
  \includegraphics[width=\columnwidth]{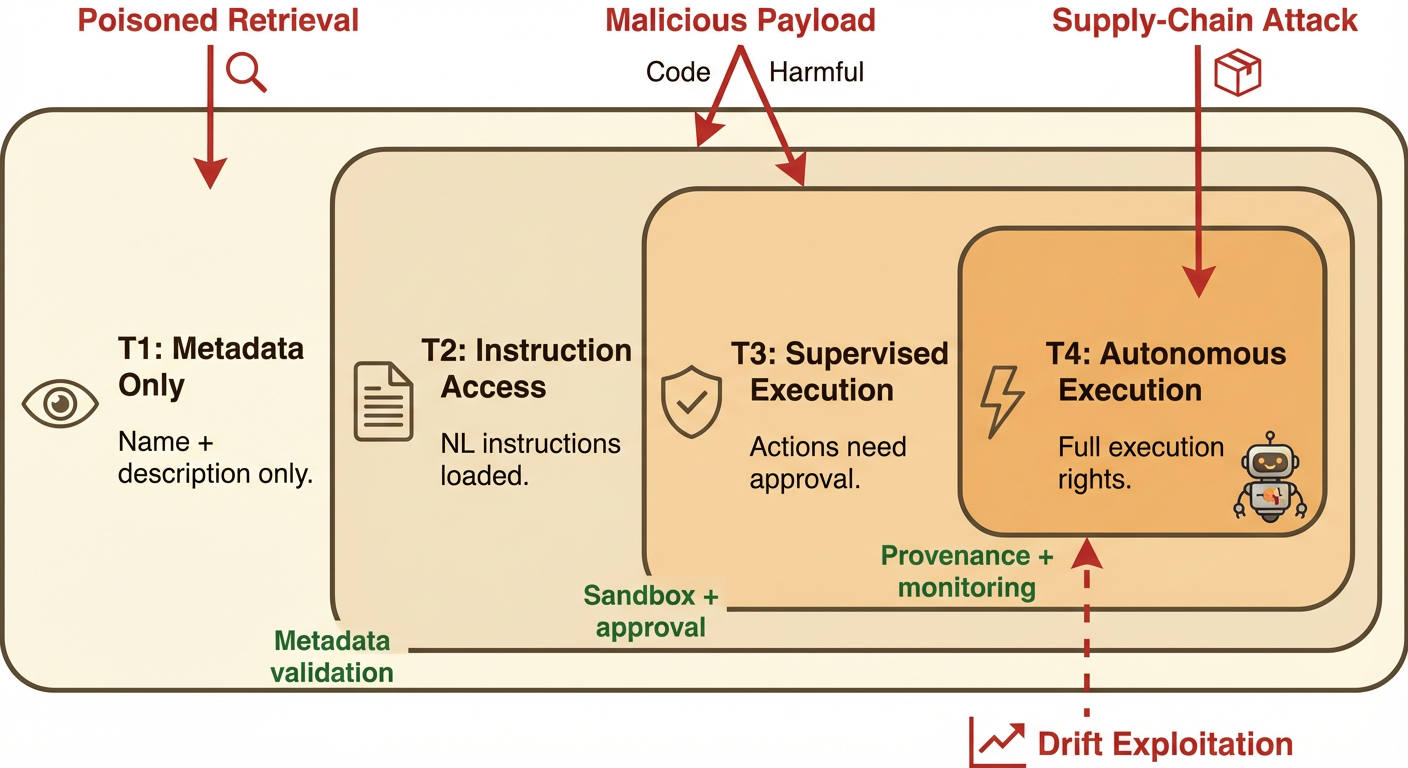}
  \caption{Trust-tiered threat model for skill governance. Four nested privilege tiers (T1--T4) form concentric security boundaries. Red arrows show attack vectors targeting different tier boundaries; green labels indicate defense mechanisms between tiers.}
  \label{fig:trust_tiers}
\end{figure}

We propose a four-tier trust model for skills. Fig.\ref{fig:trust_tiers} depicts the nested trust boundaries alongside attack vectors and defense mechanisms.

\begin{itemize}[nosep,leftmargin=*]
  \item \textbf{Tier-1}(\textit{metadata only}): The agent sees only the skill name and description. No instructions or code are loaded. This tier supports skill discovery without execution risk.
  \item \textbf{Tier-2} (\textit{instruction access}): The agent loads the skill's natural-language instructions into its context window. The instructions may influence the agent's reasoning. However, Tier-2 provides meaningful isolation only when the runtime enforces a read-only mode during instruction loading, with tool execution gated behind a separate approval channel. Without architectural separation between reasoning and action, Tier-2 instructions can indirectly induce tool invocations through the agent's standard decision loop, effectively degrading to Tier-3.
  \item \textbf{Tier-3} (\textit{supervised execution}): The skill can execute actions (tool calls, code execution) but each action requires user approval or runs within a constrained sandbox.
  \item \textbf{Tier-4} (\textit{autonomous execution}): The skill executes without per-action approval, subject to pre-configured permission boundaries and monitoring.
\end{itemize}

Production systems should default to Tier-1 for untrusted skills and require explicit trust escalation, backed by provenance verification, for higher tiers. The trust tier should be \emph{sticky}: once a skill demonstrates reliable behavior at Tier-3 over multiple invocations, it may be promoted to Tier-4, but a single safety violation should trigger demotion.

\para{Privilege escalation} The trust tier model must also guard against escalation: a Tier-1 skill's metadata could include instructions designed to trick the agent into loading it at a higher tier. Tier transitions should be enforced by the runtime, not by skill-provided metadata. Cross-referencing with prompt injection attacks~\cite{greshake2023promptinjection}, Tier-2 instruction access is particularly vulnerable when the loaded instructions contain embedded directives that cause the agent to invoke tools or escalate the skill's own privileges.

\subsection{Sandboxing and Permission Boundaries}
\label{sec:security:sandboxing}

Code skills (Pattern-2) require sandboxed execution environments that limit access to the file system, network, and system resources. Container-based sandboxes (e.g., Docker) and WebAssembly runtimes provide isolation with varying performance overhead. The key design question is granularity: should sandboxing be per-skill (each skill runs in its own sandbox), per-session (all skills in a session share a sandbox), or per-tier (sandboxing varies by trust level)?

Natural-language skills (Pattern-5) present a different sandboxing challenge: the ``execution environment'' is the agent's context window, and the ``sandbox'' is the instruction-following boundary. Prompt injection attacks~\cite{greshake2023promptinjection} demonstrate that this boundary is permeable. Architectural mitigations include separating skill instructions from user data, using structured input/output schemas, and employing output filtering to detect unauthorized actions.

\subsection{Skill Supply-Chain Governance}
\label{sec:security:supplychain}

Marketplace-distributed skills (Pattern-7) face supply-chain risks analogous to those in package management ecosystems. We recommend four governance mechanisms:

\para{Provenance signing} Each skill package includes a cryptographic signature from its author, enabling verification of authorship and integrity. This mirrors code signing in traditional software distribution.

\para{Dependency auditing} Skills may depend on other skills, tools, or external services. A dependency graph should be maintained and audited for known vulnerabilities, similar to dependency scanning in npm or pip.

\para{Continuous monitoring} Even after initial vetting, skills should be monitored for behavioral anomalies during execution. Unexpected tool calls, excessive resource consumption, or access to out-of-scope resources should trigger alerts and potential demotion to a lower trust tier.

\para{Version pinning} Skill consumers should pin to specific versions rather than tracking ``latest,'' to prevent a compromised update from automatically propagating to all consumers.

\subsection{Pattern-Specific Risk Matrix}
\label{sec:security:riskmatrix}

Different design patterns expose different attack surfaces. Table~\ref{tab:risk_matrix} maps each pattern to its primary risks and recommended mitigations.

\begin{table*}[t]
\centering
\caption{Pattern-specific security risk matrix.}
\label{tab:risk_matrix}
\resizebox{\textwidth}{!}{%
\begin{tabular}{c|c|L{5.5cm}|L{5cm}|c}
\toprule
\textbf{\#} & \textbf{Pattern} & 
\multicolumn{1}{c|}{\textbf{Primary Risks}} & 
\multicolumn{1}{c}{\textbf{Recommended Mitigations}} & \textbf{Severity} \\
\midrule
\rowcolor{ptwo!30}
1 & Metadata progressive disclosure & Metadata poisoning; misleading descriptions & Metadata schema validation; human review for high-privilege skills & Medium \\
2 & Code-as-skill & Code injection; sandbox escape; dependency vulnerabilities & Container sandboxing; static analysis; dependency scanning & High \\
\rowcolor{ptwo!30}
3 & Workflow enforcement & Rule bypass via prompt injection; overly rigid constraints & Input sanitization; rule integrity verification & Medium \\
4 & Self-evolving libraries & Poisoned distillation; skill drift; quality degradation & Human-in-the-loop verification; regression testing; anomaly detection & High \\
\rowcolor{ptwo!30}
5 & Hybrid NL+code macros & Boundary ambiguity exploitation; conflicting instructions & Clear NL/code separation; instruction priority rules & Medium \\
6 & Meta-skills & Recursive error amplification; adversarial skill generation & Generation caps; quality gates at each iteration; diversity checks & High \\
\rowcolor{ptwo!30}
7 & Marketplace distribution & Supply-chain attacks; malicious packages; version tampering & Provenance signing; continuous monitoring; version pinning & Critical \\
\midrule
--- & Confused deputy (cross-cutting) & Environmental injection coerces misuse of privileged skills (affects P1, P2, P5) & Data-flow tracking; capability confinement; input/output separation & High \\
\rowcolor{ptwo!30}
--- & $C$-poisoning (cross-cutting) & Adversarial inputs cause inappropriate skill activation (affects P1, P4) & Adversarial testing of applicability predicates; input validation & Medium \\
\bottomrule
\end{tabular}%
}
\end{table*}

\subsection{Case Study: ClawHavoc Supply-Chain Attack}
\label{sec:security:casestudy}

The ClawHavoc campaign against OpenClaw's ClawHub skill registry~\cite{clawhavoc2026} provides the first large-scale empirical evidence of skill supply-chain exploitation, concretizing every threat category in our model and revealing the severity of real-world consequences.

\para{Scale and attack surface} Within weeks of ClawHub's launch, security researchers identified {1{,}184 malicious skills across the registry~\cite{clawhavoc2026}, while a separate Snyk audit found that \textbf{36.8\%} of all published skills contained at least one security flaw. The campaign involved 12 publisher accounts, with a single account responsible for 677 packages (57\% of all malicious listings), while the platform's most-downloaded skill (``What Would Elon Do'') contained 9 vulnerabilities including 2 critical ones, with its ranking artificially inflated through 4{,}000 faked downloads~\cite{clawhavoc2026}. VirusTotal's analysis of over 3{,}016 ClawHub skills confirmed that hundreds exhibited malicious characteristics~\cite{virustotal2026openclaw}. Separately, a Snyk audit found that 283 of 3{,}984 skills (7.1\%) exposed sensitive credentials in plaintext through LLM context windows and output logs. The attack surface was global: over 135{,}000 exposed OpenClaw instances were detected across 82 countries.

\para{Severity of credential and asset theft} The consequences of malicious skill execution were not theoretical. The primary payload, Atomic macOS Stealer (AMOS), systematically harvested: (i) \textit{LLM API keys} from \texttt{.env} files and OpenClaw configuration, enabling billing fraud and model abuse; (ii)~\textit{cryptocurrency wallet keys} across \textit{60+ wallet types} including Phantom, MetaMask, and Exodus, enabling irreversible asset theft; (iii)~browser-stored \textit{passwords, credit card numbers, and autofill data} across Chrome, Safari, Firefox, Brave, and Edge; (iv)~\textit{SSH keys and Keychain credentials}, granting persistent access to production infrastructure; and (v)~Telegram sessions and local files from Desktop and Documents directories. Windows-targeted payloads delivered VMProtect-packed infostealers via password-protected archives, and 91\% of malicious skills included prompt injection payloads that weaponized the agent itself as an accomplice, attacking both humans and AI simultaneously. Belgium's Centre for Cybersecurity and China's MIIT issued emergency advisories, while multiple South Korean technology companies blocked OpenClaw entirely.

\para{Attack vector analysis through our pattern taxonomy} The ClawHavoc campaign instantiates multiple threat categories from \S\ref{sec:security:threats}:

\begin{itemize}[nosep,leftmargin=1em]
  \item \textit{Poisoned skill retrieval}: Attackers cloned popular legitimate skills under near-identical names, exploiting Pattern-1's metadata-driven discovery to rank malicious versions alongside or above originals.
  \item \textit{Malicious skill payloads}: Skills included reverse shells, credential-exfiltration webhooks, and social-engineering ``setup'' instructions that told users to run \texttt{curl~|~bash} pipelines. These exploit Pattern-2's code execution and Pattern-5's ambiguity at the NL/code boundary.
  \item \textit{Confused deputy}: Prompt injection payloads in skill documentation coerced the agent into executing malicious commands using its legitimate tool access, bypassing any skill-level trust check.
  \item \textit{$C$-poisoning}: Overbroad skill descriptions ensured malicious skills activated across broad task categories (crypto, productivity, automation), maximizing the attack surface through Pattern-1 metadata manipulation.
\end{itemize}

\para{Pattern-specific impact gradient} The severity varies by pattern. \textit{Pattern-7 (marketplace)} is the most directly impacted: it is the distribution channel through which all attacks propagated, and the ClawHub registry's initial absence of provenance signing, dependency auditing, or automated scanning enabled the 36.8\% malicious-skill rate. \textit{Pattern-2 (code-as-skill)} is the main execution vector: OpenClaw skills run code with the agent's full system permissions, so one malicious skill can access local credentials such as API keys, wallets, browser vaults, and SSH keys. \textit{Pattern-1 (metadata)} is the discovery vector: poisoned metadata enabled ranking manipulation and name-squatting. \textit{Pattern-5 (hybrid NL+code)} is exploited through documentation-as-attack-surface: skill README files contained the actual social-engineering payload. \textit{Pattern-3 (workflow enforcement)} is less exposed because hard-gated execution sequences constrain the agent's action space. For example, a mandated test-before-deploy workflow is harder to bypass through prompt injection alone. \textit{Pattern-4 (self-evolving)} presents a latent risk: if an agent's self-generated skill library ingests a malicious community skill as a template, the poison propagates through the agent's own generation loop.

\para{Governance response} OpenClaw's initial response was a partnership with VirusTotal~\cite{virustotal2026openclaw} to scan published skills using SHA-256 fingerprints, Code Insight (LLM-based behavioral analysis), and daily re-scans. This matches several mechanisms in \S\ref{sec:security:supplychain}: automated provenance checks, behavioral anomaly detection, and blocking known-bad versions by hash. OpenClaw also noted that VirusTotal scanning is ``not a silver bullet''.

\para{Why traditional scanners fail: tuple-level analysis}
The limitations of traditional malware scanners become evident when viewed through our formal definition $S = (C, \pi, T, R)$ introduced in \S\ref{sec:def:formal}. Each component of the tuple exposes a distinct attack surface, yet conventional security tools cover only a small subset of them:

\begin{itemize}[nosep,leftmargin=1em]
    \item \textit{$R$ (interface)}: The callable interface (skill name, description, and parameter schema) serves as the first point of contact. Name squatting, misleading descriptions, and inflated download counts can manipulate $R$ and distort \emph{discovery}. However, VirusTotal does not assess the semantics of skill metadata.
  \item \textit{$C$ (applicability condition)}: Overbroad applicability predicates that return $C(o,g)=1$ for maximally many contexts increase the blast radius of a malicious skill. No current scanner audits whether a skill's activation scope is proportionate to its stated purpose.
  \item \textit{$\pi$ (policy)}: The policy component is where most attacks arise, yet $\pi$ in agentic skills is inherently \emph{heterogeneous}: it may contain executable code (amenable to static analysis), natural-language instructions (largely invisible to binary scanners), or both. For example, a \texttt{curl} command hidden within a “stock tracking” skill’s setup instructions, or a prompt-injection directive (e.g., ``ignore all previous safety guidelines'') embedded in the NL policy, can exfiltrate \texttt{.env} files and API keys to external servers. VirusTotal, which is designed to detect binary malware signatures, labels such payloads as \texttt{Benign} because they appear as syntactically valid text or harmless shell commands when analyzed in isolation.
  \item \textit{$T$ (termination condition)}: A malicious $T$ can terminate the skill prematurely to evade logging (exfiltrate-then-exit-cleanly), or fail to terminate to enable persistent background access. Neither behavior triggers traditional antivirus heuristics.
\end{itemize}

\para{Complementary skill auditing} Recognizing this gap, the community has developed \emph{skill-native auditing tools} that operate at the tuple level rather than the binary level. Agent Skills Guard~\cite{agentskillsguard2026} and SkillGuard~\cite{skillguard2026} exemplify a three-layer detection architecture mapped to our formalization:

\begin{itemize}[nosep,leftmargin=1em]
\item \textit{Rule engine / AST analysis} (auditing $\pi_{\text{code}}$ and $R$): Pattern rules and Abstract Syntax Tree analysis flag risky constructs in the executable part of $\pi$ (e.g., shell execution, \texttt{eval()}, reverse shells, credential access, destructive operations) and can also catch hardcoded secrets exposed via $R$'s metadata. This layer runs locally with low overhead, incurs no per-call cost, and can cover a broad set of attack patterns across multiple languages.
\item \textit{LLM semantic analysis} (auditing $\pi_{\text{NL}}$ and $C$): An LLM reviews the natural-language part of $\pi$ for hidden intent (prompt injection~\cite{liu2023jailbreaking}, social-engineering directives, or instructions conflicting with the stated purpose) and checks whether $C$'s activation scope is appropriate. This catches attacks that rule-based or binary scanning misses, such as benign-looking NL instructions that steer the agent to exfiltrate data using otherwise legitimate tools.
\item \textit{Reputation scoring} (aggregating across $C$, $\pi$, $T$, $R$): Signals from the layers are combined into a 0-100 reputation score. The tool uses threshold bands (e.g., above 80 as ``safe'' and below 30 as ``malicious''). The authors report a controlled evaluation on 39~test cases, including 4~adversarial samples that VirusTotal marked as \texttt{Benign}, and no false positives on legitimate skills in that set~\cite{agentskillsguard2026}.
\end{itemize}

These tools are implemented as agent skills (Pattern-2) and can be installed directly in the OpenClaw environment. This means an agent can use one skill to audit others. In practice, it shows that the skill abstraction can also encode governance checks within the $(C, \pi, T, R)$ framework.
The ClawHavoc case demonstrates that skill marketplace governance requires \emph{defense in depth}: binary scanning (VirusTotal) catches commodity malware targeting $\pi_{\text{code}}$, skill-native auditing catches NL-level attacks on $\pi_{\text{NL}}$ and $C$, and runtime behavioral monitoring is needed to detect attacks on $T$ and context-dependent exploits that only manifest during execution. No single layer suffices; the tuple-level decomposition provides the conceptual framework for understanding which defenses cover which attack surfaces.

\section{Evaluating Agentic Skills}
\label{sec:evaluation}

We evaluate the utility of agentic skills through a five-dimensional framework and map existing benchmarks to measurable skill properties.

\subsection{Evaluation Dimensions}
\label{sec:evaluation:dimensions}

\para{Correctness}
Correctness measures whether a skill achieves its intended outcome. Evaluation relies on ground-truth annotations or deterministic verifiers. For code skills, unit tests provide direct verification, while for web interaction skills, environment state comparison (e.g., verifying whether a form was submitted correctly) serves as a practical proxy.

\para{Robustness}
Robustness captures a skill’s reliability under input variations, environment perturbations, and edge cases. A robust skill maintains consistent performance when confronted with minor deviations from the training distribution, such as handling both legacy and updated UI layouts.

\para{Efficiency}
Efficiency characterizes the resource cost of executing a skill. Relevant metrics include token consumption (for natural-language skills), wall-clock time, number of tool calls, and API costs. Efficiency directly affects deployment cost and composability, as inefficient sub-skills slow downstream workflows.

\para{Generalization}
Generalization evaluates whether a skill transfers to unseen tasks or domains. This dimension is challenging to measure because it requires out-of-distribution evaluation. Benchmarks such as cross-website generalization in Mind2Web~\cite{deng2023mind2web} and cross-application evaluation in OSWorld~\cite{xie2024osworld} provide partial evidence.

\para{Safety}
Safety assesses whether a skill avoids harmful actions, respects permission boundaries, and handles failures gracefully. Evaluation commonly involves adversarial testing, red-teaming, and runtime monitoring for unauthorized or unsafe behaviors.

\subsection{Deterministic Evaluation Harnesses}
\label{sec:evaluation:harnesses}

Human evaluation of agent skills does not scale. We advocate for \emph{deterministic evaluation harnesses}: benchmark environments where success is measured automatically by checking environment state against expected outcomes. This approach provides low-cost reproducible evaluation that can be integrated into skill development pipelines.

The key design principle is \emph{outcome-based verification}: rather than judging the quality of intermediate reasoning or the elegance of the skill's approach, the harness checks whether the intended outcome was achieved. This aligns with the pragmatic nature of skills as procedural modules valued for their effects, not their form.

\subsection{Benchmark-to-Skill Mapping}
\label{sec:evaluation:benchmarks}

Table~\ref{tab:benchmark_mapping} maps major agent benchmarks to the skill dimensions they assess. No single benchmark covers all dimensions; a comprehensive skill evaluation requires combining multiple benchmarks.

\begin{table*}[t]
\centering
\caption{Benchmark-to-skill-dimension mapping. \cmark\ = primary assessment; \pmark\ = partial assessment; empty = not assessed.}
\label{tab:benchmark_mapping}
\resizebox{\textwidth}{!}{%
\begin{tabular}{c|c|c|c|c|c|c|l}
\toprule
\textbf{Benchmark} & \textbf{Environment} & \textbf{Correctness} & \textbf{Robustness} & \textbf{Efficiency} & \textbf{Generalization} & \textbf{Safety} & \textbf{Skill Scope Assessed} \\
\midrule
SkillsBench~\cite{skillsbench2025} & Multi & \cmark & \pmark & \cmark & \cmark & & Skill utility, composition, domain \\
WebArena~\cite{zhou2024webarena} & Web & \cmark & \pmark & \pmark & & & Web navigation, UI grounding \\
Mind2Web~\cite{deng2023mind2web} & Web & \cmark & & & \cmark & & Cross-site generalization \\
OSWorld~\cite{xie2024osworld} & Desktop & \cmark & \pmark & & \cmark & & Multi-application workflows \\
SWE-bench~\cite{jimenez2024swebench} & SWE & \cmark & & \pmark & \pmark & & Code understanding, patch generation \\
GAIA~\cite{mialon2023gaia} & Multi & \cmark & & & \cmark & & General assistant capability \\
AgentBench~\cite{liu2023agentbench} & Multi & \cmark & \pmark & \pmark & \cmark & & Cross-environment performance \\
AndroidWorld~\cite{rawles2024androidworld} & Mobile & \cmark & \pmark & & & & Mobile interaction skills \\
\bottomrule
\end{tabular}%
}
\end{table*}

\subsection{Anchor Case Study: SkillsBench}
\label{sec:evaluation:skillsbench}

The SkillsBench benchmark~\cite{skillsbench2025} provides the most direct evidence to date for the value of curated skills. We note that the quantitative findings in this subsection derive primarily from a single, non-peer-reviewed benchmark. While the scale (86~tasks, 7{,}308~trajectories) and methodological rigor of SkillsBench provide useful evidence, independent replication across additional benchmarks is needed to confirm these patterns. The benchmark evaluates 86~tasks across 11~domains (healthcare, manufacturing, cybersecurity, natural science, energy, finance, office work, media, robotics, mathematics, and software engineering) using 7~agent-model configurations over 7{,}308~trajectories. Each task is assessed under three conditions: no~skills, curated~skills, and self-generated~skills, with deterministic verifiers ensuring objective evaluation.

\para{Curated skills provide substantial, quantifiable improvement} Across all configurations, curated skills raise the average pass rate by \textit{16.2}~percentage points (from 24.3\% to 40.6\%). The effect varies dramatically by domain: healthcare sees +51.9\,pp, manufacturing +41.9\,pp, and cybersecurity +23.2\,pp, while software engineering gains only +4.5\,pp and mathematics +6.0\,pp. This domain variance is consistent with the hypothesis that skills help most where the base model's pretraining data provides insufficient procedural grounding, which is directly relevant to the scope axis of our taxonomy (\S\ref{sec:taxonomy:scope}). Domain variance may also reflect confounders including task construction, verifier strictness, and skill authoring quality differences across domains.

\para{Self-generated skills provide no benefit} Self-generated skills average $-$1.3\,pp relative to the no-skills baseline, suggesting that models cannot yet reliably author the procedural knowledge they benefit from consuming in open-ended settings. Only one configuration (Claude Opus~4.6) showed a modest +1.4\,pp, while Codex + GPT-5.2 degraded by $-$5.6\,pp. This finding is consistent with the quality concerns raised for self-evolving libraries (Pattern-4, \S\ref{sec:patterns:selfevolving}).

\para{Skill quantity and complexity matter} Focused skills with 2--3~modules yield optimal improvement (+18.6\,pp), while 4+ skills show diminishing returns (+5.9\,pp).  ``Detailed'' skills (moderate-length, focused guidance) improve by +18.8\,pp, whereas ``comprehensive'' skills (exhaustive documentation) \emph{degrade} performance by $-$2.9\,pp. This pattern is consistent with Pattern-1 (metadata-driven progressive disclosure, \S\ref{sec:patterns:metadata}): loading focused procedural instructions outperforms loading comprehensive reference material.

\para{Skills as compute equalizers} Smaller models equipped with curated skills can match or exceed larger models without skills. Claude Haiku~4.5 with skills (27.7\%) outperforms Claude Opus~4.5 without skills (22.0\%), suggesting that skill libraries may serve as a practical cost-reduction mechanism.

\para{Negative-delta tasks} 16 of 84~tasks show performance degradation with skills, with the worst case ($-$39.3\,pp) occurring for tasks where the base model already performs well and skills introduce conflicting guidance. This highlights the importance of the applicability condition $C$ in our formalization~(\S\ref{sec:def:formal}): a skill should activate only when its procedural knowledge is beneficial.

We present these as interpretive hypotheses grounded in the benchmark data, not causal conclusions; Voyager's self-verification success rate and AgentBench~\cite{liu2023agentbench} cross-environment results provide partial corroboration from independent sources, even if they do not directly measure the curated-vs-self-generated comparison. These findings underscore the importance of distinguishing between \emph{skill availability} (having relevant skills) and \emph{skill quality} (having skills that actually help). The skill lifecycle model (\S\ref{sec:lifecycle}) addresses both: discovery and storage ensure availability, while practice, evaluation, and update ensure quality.

\section{Discussion and Limitations}
\label{sec:discussion}

\subsection{Cross-Cutting Observations}
\label{sec:discussion:crosscutting}

Several patterns emerge from the systematization that are not visible from any individual system.

\para{Representation--governance coupling} More formal skill representations admit stronger governance. Code skills (Pattern-2) support static analysis, unit testing, and sandboxed execution; natural-language skills resist all three. This creates a tension: the representations that are easiest to author (NL) are hardest to govern, while those amenable to formal verification (code, policy) require specialized authoring expertise. No existing system resolves this tension fully; hybrid representations (Pattern-5) attempt a compromise but introduce boundary ambiguity.

\para{Sparsity of the design space} Table~\ref{tab:taxonomy_master} reveals that most systems cluster in a narrow region: code-as-skill representation with self-evolving library patterns in game or SWE environments. Large regions of the representation $\times$ scope $\times$ pattern space remain unexplored, particularly policy-based skills with marketplace distribution and NL skills with formal workflow enforcement. These unexplored regions represent both opportunity and risk: they may be inherently difficult (explaining the sparsity) or simply underexplored.

\para{Marketplace growth outpaces governance} TThe OpenClaw experience (\S\ref{sec:security:casestudy}) shows that when skill ecosystems grow quickly, governance mechanisms can lag behind. ClawHub's 36.8\% malicious-skill rate at its peak is orders of magnitude worse than the point-in-time malware rates observed in mature package registries such as npm, reflecting the absence of even basic supply-chain protections (package signing, automated scanning, reputation scoring) that took traditional package ecosystems years to develop. The subsequent VirusTotal partnership reduced the threat surface but was reactive; proactive governance (pre-publication scanning, behavioral sandboxing, capability confinement) is necessary for Pattern-7 systems that distribute skills with full system access. This observation reinforces the Pattern-3 advantage in our taxonomy: workflow enforcement, which constrains execution sequences \emph{before} skills run, is inherently more resilient to supply-chain compromise than patterns that grant broad execution permissions and attempt to detect misuse \emph{after} the fact.

\para{The curation-scalability tradeoff} The SkillsBench evidence (\S\ref{sec:evaluation:skillsbench}) quantifies a fundamental tradeoff: curated skills improve pass rates by +16.2\,pp on average, while self-generated skills degrade them by $-$1.3\,pp. Self-evolving libraries (Pattern-4) are the most scalable acquisition mechanism but produce skills that hurt performance; human curation yields the most reliable skills but does not scale. SkillsBench further shows that focused skills with 2--3~modules outperform comprehensive documentation, suggesting that the quality problem is not just accuracy but also \emph{conciseness}: effective skills must distill procedural knowledge rather than dump reference material. The tension is not absolute. Verification-gated self-generation (Voyager~\cite{wang2023voyager}, Eureka~\cite{ma2024eureka}) succeeds in constrained environments with deterministic execution feedback; the SkillsBench evidence indicates this success does not yet generalize to open-ended, multi-domain settings without execution-verified practice loops. Closing the tradeoff likely requires combining autonomous generation with automated verification pipelines and length-constrained distillation.

\subsection{Limitations of This Systematization}
\label{sec:discussion:limitations}

\para{Corpus recency} The LLM agent skill ecosystem is recent: the majority of systems analyzed were published in 2023--2024. While we ground the skill abstraction in decades of cognitive science and RL, the LLM-specific literature may be too nascent for patterns identified here to be stable. Taxonomies may require revision as the field matures.

\para{Corpus coverage} Our analysis examines 24 systems from a retained set of 65 papers. Despite systematic search procedures (\S\ref{sec:methodology}), we may miss relevant work, particularly from industry systems with limited public documentation, non-English-language publications, and concurrent preprints.

\para{Taxonomy validation} The seven design patterns were derived bottom-up from the analyzed systems but have not been validated through external expert surveys or formal concept analysis. The non-exclusivity of patterns (systems combine multiple patterns) complicates categorical analysis and may limit the taxonomy's discriminative power for future systems that combine patterns in novel ways.

\para{Production and safety coverage} Our corpus emphasizes research systems with published evaluations. Several production frameworks (e.g., LangChain/LangGraph for skill composition, DSPy for declarative skill compilation) and safety-focused benchmarks (e.g., AgentHarm, InjectAgent) are relevant but under-represented in our analysis due to limited peer-reviewed documentation. We focus on systems with sufficient published detail for rigorous classification.

\para{Benchmark reliance} Our evaluation analysis relies on published benchmark results, which may not reflect real-world skill utility. Production deployments involve longer time horizons, messier environments, and adversarial conditions not captured by existing benchmarks.

\section{Open Problems and Research Roadmap}
\label{sec:openproblems}

Existing skill-based agents still expose several unresolved tensions that limit reliable deployment at scale. We highlight several directions.

\subsection{Verified Autonomous Skill Generation}
\label{sec:open:verified}

A central tension revealed throughout our analysis is the trade-off between scalability and reliability in skill construction. Systems that allow skills to evolve autonomously can expand capability libraries rapidly, yet empirical evidence shows that automatically generated skills may occasionally degrade downstream performance. In contrast, human-curated skills remain more dependable but introduce a clear scalability bottleneck, as manual validation cannot keep pace with growing agent deployments. 

This shows that the key obstacle is no longer skill generation itself, but verification at the point of admission into the skill library. A promising direction is to treat skills similarly to software artifacts in continuous integration pipelines: newly generated skills would be evaluated against held-out task distributions before becoming reusable components. For code-centric skills, formal or semi-formal verification techniques may provide guarantees about behavior, while natural-language or hybrid skills likely require behavioral testing and regression-style evaluation. Progress in this area would enable self-evolving skill libraries that improve over time without accumulating hidden performance regressions.

\subsection{Unsupervised Skill Discovery}
\label{sec:open:discovery}

Another limitation concerns how new skills are discovered in the first place. Although many existing systems advertise autonomous learning, most still rely heavily on external scaffolding such as predefined curricula, demonstrations, or explicit reward signals. Our lifecycle survey shows that fully autonomous discovery remains rare: even systems designed for exploration typically depend on some form of human guidance to define progress.

Achieving open-ended capability growth therefore requires moving beyond supervised discovery. One possible path is adapting unsupervised skill discovery techniques from reinforcement learning to LLM-based agents, allowing reusable behaviors to emerge directly from interaction traces. Signals such as repeated trajectory patterns, attention regularities, or recurring subgoal structures may serve as implicit indicators of skill boundaries. An agent capable of extracting reusable competencies solely from its own experience would fundamentally change how agent capabilities scale, shifting learning from instruction-driven expansion toward self-organizing behavior.

\subsection{Formal Verification Across Representations}
\label{sec:open:verification}

A practical governance challenge arises from the diversity of skill representations. Skills expressed as executable code benefit from decades of software assurance techniques, including testing, static analysis, and sandboxing. In practice, however, many deployed skill libraries rely heavily on natural-language or policy-style skills because they are easier to author and distribute. Unfortunately, these representations are significantly harder to audit rigorously, creating a mismatch between expressive convenience and verifiability.

This gap becomes particularly visible in safety-sensitive deployments, where auditing requirements extend beyond simple correctness checks. Emerging approaches suggest combining multiple lightweight verification layers: rule-based analysis for executable components, semantic inspection for language-based policies, and reputation or behavioral monitoring across executions. The longer-term challenge is moving from static, pre-deployment inspection toward runtime verification capable of detecting context-dependent failures or delayed activation attacks that only appear under specific environmental conditions.

\subsection{Robustness Under Environmental Drift}
\label{sec:open:drift}

Even correctly implemented skills may fail over time as their operating environments evolve. Changes in APIs, tools, data formats, or surrounding workflows can gradually invalidate assumptions embedded in a skill, producing unintended behavior without any modification to the skill itself. This form of environmental drift creates an attack surface that operates indirectly: adversaries can manipulate external conditions rather than the skill artifact.

Despite its practical importance, proactive drift detection remains largely absent from current systems. Future work may focus on continuous monitoring mechanisms that track execution statistics, detect deviations from historical behavior, and correlate failures with environmental change signals. Such systems would treat skills as living components requiring maintenance, enabling automatic adaptation or retirement once reliability deteriorates. Addressing drift will likely become essential as agents transition from experimental settings to long-lived production deployments.

\subsection{Governance Economics and Liability}
\label{sec:open:governance}

Finally, the emergence of marketplace-style skill distribution introduces economic and governance questions that remain largely unexplored. Open skill ecosystems create strong incentives for contribution and innovation, but simultaneously expand the supply-chain attack surface. Our survey indicates that existing platforms rarely provide clear mechanisms for assigning responsibility when third-party skills cause harm, nor do they offer credible certification processes that align incentives with reliability.

Progress here requires integrating technical and economic design. Liability models must clarify responsibility among skill authors, platform operators, and users, while certification mechanisms should reward dependable skills and discourage risky ones. Understanding these dynamics may require agent-based economic modeling alongside empirical platform studies. As skill marketplaces mature, governance frameworks that combine accountability, certification, and incentive alignment will likely become as important as technical advances themselves.

\section{Conclusion}
\label{sec:conclusion}

Agentic skills are reusable procedural modules for LLM agents. We structure the design space, analyze security risks, and show that skill quality critically affects agent performance. We close by outlining open challenges in discovery, verification, and governance for reliable skill-based agents.

\balance
\bibliographystyle{IEEEtran}
\bibliography{references}

\end{document}